\documentclass[twocolumn]{aastex63}
\hypersetup{linkcolor=blue,citecolor=blue,filecolor=cyan,urlcolor=cyan}
\usepackage{comment}
\graphicspath{{../}{./}{figures/}}

\accepted{24 January 2021}
\submitjournal{ApJ}

\shorttitle{MPs in AGB stars of Galactic GCs}
\shortauthors{Lagioia et al.}
\newcommand{\msun}{$M_{\odot}$}
\newcommand{\ngc}{NGC\,}
\newcommand{\hst}{\textit{HST}}

\newcommand{\feh}{\rm [Fe/H]}
\newcommand{\afe}{\rm [$\alpha$/Fe]}
\newcommand{\teff}{$T_{\rm eff}$}
\newcommand{\ome}{$\rm \omega\,Cen$}
\newcolumntype{C}{>{$}c<{$}}

\defcitealias{milone17}{M17}
\defcitealias{milone18b}{M18B}

\begin{document}

\title{Multiple stellar populations in Asymptotic Giant Branch stars of Galactic Globular Clusters} %\footnote{Released on}}

\correspondingauthor{Edoardo Lagioia}
\email{edoardo.lagioia@unipd.it}

\author[0000-0003-1713-0082]{E. P. Lagioia} 
\affiliation{Dipartimento di Fisica e Astronomia ``Galileo Galilei'', Universit\`{a} di Padova, Vicolo dell'Osservatorio 3, 35122, Padova, Italy}
\affiliation{Istituto Nazionale di Astrofisica - Osservatorio Astronomico di Padova, Vicolo dell'Osservatorio, 5, 35122, Padova, Italy}

\author[0000-0001-7506-930X]{A. P. Milone}
\affiliation{Dipartimento di Fisica e Astronomia ``Galileo Galilei'', Universit\`{a} di Padova, Vicolo dell'Osservatorio 3, 35122, Padova, Italy}
\affiliation{Istituto Nazionale di Astrofisica - Osservatorio Astronomico di Padova, Vicolo dell'Osservatorio, 5, 35122, Padova, Italy}

\author[0000-0002-1276-5487]{A. F. Marino}
\affiliation{Istituto Nazionale di Astrofisica - Osservatorio Astronomico di Arcetri , Largo Enrico Fermi, 5, 50125, Firenze, Italy}

\author[0000-0002-1128-098X]{M. Tailo}
\affiliation{Dipartimento di Fisica e Astronomia ``Galileo Galilei'', Universit\`{a} di Padova, Vicolo dell'Osservatorio 3, 35122, Padova, Italy}

\author[0000-0002-7093-7355]{A. Renzini}
\affiliation{Istituto Nazionale di Astrofisica - Osservatorio Astronomico di Padova, Vicolo dell'Osservatorio, 5, 35122, Padova, Italy}

\author[0000-0003-1757-6666]{M. Carlos}
\affiliation{Dipartimento di Fisica e Astronomia ``Galileo Galilei'', Universit\`{a} di Padova, Vicolo dell'Osservatorio 3, 35122, Padova, Italy}

\author[0000-0002-7690-7683]{G. Cordoni}
\affiliation{Dipartimento di Fisica e Astronomia ``Galileo Galilei'', Universit\`{a} di Padova, Vicolo dell'Osservatorio 3, 35122, Padova, Italy}

\author[0000-0001-8415-8531]{E. Dondoglio}
\affiliation{Dipartimento di Fisica e Astronomia ``Galileo Galilei'', Universit\`{a} di Padova, Vicolo dell'Osservatorio 3, 35122, Padova, Italy}

\author[0000-0002-1562-7557]{S. Jang}
\affiliation{Dipartimento di Fisica e Astronomia ``Galileo Galilei'', Universit\`{a} di Padova, Vicolo dell'Osservatorio 3, 35122, Padova, Italy}

\author[0000-0002-3625-6951]{A. Karakas}
\affiliation{School of Physics and Astronomy, Monash University, VIC 3800, Australia; ARC Centre of Excellence for All Sky Astrophysics in 3 Dimensions (ASTRO 3D), Australia}

\author[0000-0002-4442-5700]{A. Dotter}
\affiliation{Harvard-Smithsonian Center for Astrophysics, Cambridge, MA 02138, USA}

\begin{abstract}
Multiple stellar populations (MPs) are a distinct characteristic of Globular
Clusters (GCs). Their general properties have been widely studied among
main sequence, red giant branch (RGB) and horizontal branch (HB) stars, but a
common framework is still missing at later evolutionary stages.

We studied the MP phenomenon along the AGB sequences in 58 GCs, observed with the
\textit{Hubble Space Telescope} in ultraviolet (UV) and optical filters. By
using UV-optical color-magnitude diagrams, we selected the AGB members
of each cluster and identified the AGB candidates of the metal-enhanced population
in type\,II GCs. We studied the photometric properties of AGB stars and 
compared them to theoretical models derived from synthetic spectra analysis.

We observe the following features: i) the spread of AGB stars in 
photometric indices sensitive to variations of light-elements and helium is
typically larger than that expected from photometric errors; ii) the fraction
of metal-enhanced stars in the AGB is lower than in the RGB in most of the
type\,II GCs; iii) the fraction of 1G stars derived from the chromosome
map of AGB stars in 15 GCs is larger than that of RGB stars; v) the AGB/HB 
frequency correlates with the average mass of the most helium-enriched population.

These findings represent a clear evidence of the presence of MPs along the AGB
of Galactic GCs and indicate that a significant fraction of helium-enriched stars, 
which have lower mass in the HB, does not evolve to the AGB phase, leaving the HB 
sequence towards higher effective temperatures, as predicted by the AGB-\textit{manqu\'e} scenario.
\end{abstract}


\section{Introduction} \label{sec:intro}
A peculiar feature of the chemical composition of stars belonging to Milky Way
Globular Clusters (GCs) is the significant variation of light-elements such as
helium, carbon (C), nitrogen (N), oxygen (O) and sodium (Na), observed only in
a tiny fraction of the Galactic bulge ($\sim 1$\%, \citeauthor{schiavon17},
\citeyear{schiavon17}) and halo field stars ($\sim 3$\%,
\citeauthor{martell11a}, \citeyear{martell11a}).

In few GCs, stars also show difference in magnesium (Mg) and
aluminum (Al) content. The observed chemical variations are not random but follow trends
defined by the anti-correlation C-N, O-Na, and Mg-Al. In particular, every GC
includes two main stellar groups: 1G, composed of stars with halo-like chemical
composition; 2G composed of stars depleted in C and O, and enhanced in N, Na
and helium with respect to 1G.  When observed in color-magnitude diagrams
(CMDs) obtained by combining ultraviolet (UV) and optical observations, 1G and
2G stars lie on distinct sequences, called multiple stellar
populations (MPs).
The current framework of MPs in Galactic GCs largely derives from the study of
the spectroscopic and photometric properties of red giant branch (RGB) stars
\citep[e.g.][]{milone17,lagioia18,marino19}. In a limited number of GCs it has
also been possible to clearly identify MPs in fainter stars, thanks to the
detection of split main sequences in CMDs
\citep[e.g.]{anderson97,dantona05,piotto07,milone13,milone15b,bellini17b}.
	
% AGB problem
The study of MPs at later evolutionary stages, is more problematic. In the case
of horizontal branch (HB) stars, both age-metallicity and helium-mass loss
degeneracy hamper a proper understanding of the impact that chemical variations
have on the color distribution of core helium-burning stars in CMDs \citep[see
e.g.][]{sandage67b,fusipecci93c,catelan01,tailo20}. This indetermination
propagates to the subsequent evolutionary phase, the asymptotic giant branch
(AGB), where the uncertain definition of the sub-populations distribution is
further amplified by statistical fluctuations due to the fast evolutionary
timescale of AGB stars \citep{greggio90}. 

A viable solution to this issue is provided by the direct spectroscopic
determination of the atmospheric abundance of the light proton-capture tracing 
MPs \citep{sneden00}. In this context, early studies found evidence of
significant difference in CN band strength between RGB and AGB stars in
\ngc5904 (M\,5) and \ngc6752, with the latter being biased toward the CN-weak
populations \citet[see][]{norris81,smith93}. The same disproportion in CN-band
distribution was also detected later in \ngc1851 AGB stars \citep{campbell12}.
More recently, works based on high-resolution determination of abundance of Na,
a light element not involved in internal mixing processes during the RGB
evolution \citep{kraft94,carretta09a}, confirmed the lack of Na-rich stars in
the AGB of some GCs.

For instance, the Na content of 20 AGB stars of \ngc6752 analyzed by
\citet{campbell13}, was found to be lower than $\sim 0.2$\,dex, corresponding
to the abundance threshold separating 1G and 2G RGB stars
\citep{carretta12,milone13}. Later on, through the analysis of new
spectroscopic data of the same sample of stars, \citet{lapenna16} confirmed a
relative lack of 2G members among \ngc6752 AGB stars, with none of them
associated to the most Na- and helium-rich cluster population
\citep{carretta12,milone13}. Interestingly, this finding seems to confirm the
theoretical scenario predicted for the evolution of the hot HB stars of
\ngc6752 \citep{villanova09,cassisi14a,tailo19}. Indeed, hot HB stars with tiny
hydrogen-rich atmospheres, usually associated to the most helium-rich GC
population(s) \citep[e.g.][]{dantona02a,milone18b}, would skip the early-AGB
phase, characterized by a progressive expansion and cooling of the HB star
envelope, and instead evolve to higher temperatures and luminosities until they
reach the white-dwarf cooling sequence. The shell helium-burning phase of these
stars is therefore called AGB-\textit{manqu\'e} \citep{greggio90}.

A confirmation of this prediction would come from the recent spectroscopic
analysis of the AGB stars of \ngc2808, a GC cluster known for its extended (and
complex) HB morphology, linked to one of the largest internal helium variations
observed among Galactic GCs \citep[$\Delta$Y$\sim0.124$, with Y = helium mass
fraction,][]{milone15b}.  Indeed, \citet{marino17} identified three groups of
AGB stars with distinct Na abundance, following the typical Na-O
anti-correlation trend observed for the cluster RGB stars, with the
distribution skewed towards lower Na RGB values \citep[see also][]{wang16},
thus implying that the extremely helium-enhanced cluster stars evolved as AGB
\textit{manqu\'e}. 

However, a similar outcome has been found in \ngc6121 (M\,4) which,
at the contrary of \ngc2808, is not populated by extremely hot HB stars
\citep{marino08,villanova12} and shows a rather small internal helium
enrichment \citep[$\Delta$Y$\sim0.01$][]{milone18b}. As in the previous case,
\citet{marino17} and \citet{wang17} found that the Na-poor and Na-rich cluster
AGB star roughly cover the lowest two thirds of the [Na/Fe] abundance range
occupied by the cluster RGB stars, thus suggesting that part of the 2G cluster
stars could not reach the AGB phase \citep{maclean16}. 

Dispersion in Mg-Al abundance has also been employed to explore MPs in the AGB
stars of GCs with different HB morphology. For instance, \citet{lapenna15a}
found no Al-rich stars among the AGB stars of \ngc6266 (M\,62), a result
indicating the total lack of of 2G AGB stars and compatible with the large
helium variation detected from main sequence stars color difference by
\citet{milone15}. Conversely, in their spectroscopic analyses,
\citet{garcia-hernandez15} and \citet{masseron19} found a similar spread in the
AGB and RGB stars of \ngc5024 (M\,53), \ngc5272 (M\,3), \ngc6205 (M\,13),
\ngc6341 (M\,92), \ngc7078 (M\,15), and \ngc7089 (M\,2). 

In principle, spectroscopic discrepancies might be reduced by claiming either
zero-point offsets in the abundances obtained from different datasets or with
different reduction techniques \citep[see e.g.][]{campbell17}, or the adoption
of arbitrary thresholds for the selection of typical abundances in 1G and 2G
AGB stars \citep{marino17}. However, an effective solution to the ``AGB
problem'', namely the unexpected high fraction of 2G stars avoiding the AGB
phase \citep{campbell13}, requires an homogeneous definition of the
observational properties of these stars in large samples of GCs. In this
regard, photometry represents an ideal tool because it provides the necessary
multiplexing capability. For instance, \citet{gratton10} adopted a statistical
approach to analyze the relation between the parameter
$R_2=\mathrm{N_{AGB}/N_{HB}}$, namely the numerical fraction of AGB-to-HB stars
in a GC \citep{caputo89}, and the minimum mass along the HB, for a sample of 21
clusters. They found that blue-extended HB-morphology clusters attain $R_2$
values smaller than the rest of GCs. This result would be consistent with the
AGB-\textit{manqu\'e} prediction.

Another approach is represented by the study of MPs through the use of filter
combinations that trace the different chemical content of stars through the
variation of their flux at different wavelengths. The two main solutions
adopted so far are the index or pseudo-color $C_{\rm F275W,F336W,F438W}$ =
$(m_{\rm F275W}-m_{\rm F336W})-(m_{\rm F336W}-m_{\rm F438W})$, sensitive to C, N, O stellar
content thanks to the specific pass-band of the F275W, F336W, and F438W
filters, available at the UV and Visual (UVIS) channel of the Wide Field
Planetary Camera 3 (WFC3) on board \textit{Hubble Space Telescope}
\citep[\hst][]{milone15a}, and $C_{\rm F336W,F438W,F814W}$ =
$(m_{\rm F336W}-m_{\rm F438W})-(m_{\rm F438W}-m_{\rm F814W})$, sensitive to the content of N,
O, and helium through the F814W filter. From the previous definitions it
follows that the latter index is not as sensitive as the former to the presence
of MPs. Its lower accuracy is, however, counterbalanced by two clear upsides:
the existence of a corresponding ground-based version $C_{UBI}$ = $(U-B)-(B-I)$
\citep{marino08,monelli13} and, as a consequence, the possibility of being used
in the post-\hst\ era \citep{lagioia19a}.

The measurement of the color spread of stars in both previous combinations,
usually referred to as width, $W$, provides a general view of the total
chemical variations in the host GCs, and can be employed for a direct
comparison of the MP properties in GCs with different age, mass, and
metallicity, as already done in extensive studies of MPs in Galactic and
extra-galactic GCs, by using the width of RGB stars \citep{milone17,lagioia19}.
In principle, therefore, the same approach can also be adopted with AGB stars.
Indeed, few recent works have clearly shown that AGB stars are spread over
color intervals comparable to those observed for RGB stars, like in the case of
NGC\,7089 \citep{milone15a}, \ngc2808 \citep{milone15b}, and \ngc6121
\citep{lardo17,marino19}. Moreover, the direct comparison between AGB and
RGB star width in any given cluster could provide a solid evidence of missing
MPs in AGB phase \citep{marino17}. 

% Chromosome maps of AGB stars
The quantity $W$, although representing a direct evidence of GC internal
chemical variations, cannot provide the necessary resolution for the study of the
detailed chemical composition of a cluster. A powerful tool has been recently
introduced to overcome this problem: the so-called chromosome map
\citep[ChM,][]{milone15a,milone15b,milone17}. This photometric diagram
correlates the spread of stars in two different color combinations, each one
sensitive to different features of the spectrum of a star. The most
comprehensive ChM compilation published so far \citep{milone17} includes all
the 57 GCs observed in the \hst\ UV Legacy Survey \citep{piotto15} and has
provided the most detailed glimpse into the complex composition of Galactic
GCs. In particular, the ChM combines the information of the
$C_{\rm F275W,F336W,F438W}$ with the large-baseline color $m_{\rm F275W}-m_{\rm F814W}$,
that is sensitive to metallicity and helium variations.  In this context, the
recent analysis by \citet{marino17} has shown that the ChM of the AGB stars of
\ngc2808 present a degree of complexity comparable with that visible in the ChM
of RGB stars \citep{milone15a}, with three distinct groups corresponding to as
many different spectroscopic abundances.  It is unfortunately not as easy to
obtain the ChM of AGB stars as for the RGB for plain evolutionary reasons.
Indeed, the AGB evolutionary timescale is about ten times faster than that of
RGB stars brighter than HB stars \citep{greggio90}.  It follows that only a
limited number of GCs can provide the necessary statistical significance for a
detailed MP study. 

All the aforementioned findings represent a clear signature that chemical
variations are also a common property of AGB stars, and offer a compelling
motivation for the extension of the MP analysis to a larger sample of clusters,
which is the purpose of the present work. For this reason, we decided to
analyze the observational properties of the AGB stars of all the clusters
observed in the \hst\ UV Legacy Survey \citep{piotto15}, in the same photometric
bands adopted for the study of MPs along the RGB \citep{milone17}.

% Resume
Section~\ref{sec:obs} of the present paper describes the methodology for the
selection of the AGB and RGB samples in each analyzed cluster, the
identification of AGB stars belonging to the \deleted{anomalous}
metal-rich population (hereafter called anomalous) in type\,II GCs,
and the measurement of the color spread, or width, of the selected sample in
specific photometric indices. Section~\ref{sec:teo} reports the computation of
appropriate evolutionary models for the comparison between observed width and
theoretical predictions. Section~\ref{sec:AGBvs} illustrates the relation
existing between the AGB and RGB width. Section~\ref{sec:AGBChM} show the
procedure for the construction of the ChM of clusters with well populated AGB
samples, and the analysis of population ratios. Section~\ref{sec:frequency}
describes the relation between the frequency of AGB stars and chemical
enrichment in GCs. Finally, Section~\ref{sec:end} provides a summary of the
analysis and a description of the contribution of this work to the MP
phenomenology in GCs.

\section{Data reduction and analysis}\label{sec:obs}
The dataset analyzed in this work comprises a total of 58 clusters, including
the 57 GCs observed in the UV Legacy Survey of Galactic GCs \citep{piotto15},
and the distant Galactic cluster \ngc2419 \citep{zennaro19}. All the 57 UV
Legacy Survey GCs have been observed in the WFC3/UVIS broad-bands F275W, F336W,
and F438W, and complemented  with the F606W and F814W observations collected
for the ACS Survey of Galactic GCs \citep{sarajedini07a}. A complete
description of the image database, exposure times, and reduction techniques is
reported in \citet{piotto15,milone18b} and \citet{nardiello18a}.  Details about
database and reduction procedure of \ngc2419 data are provided in the recent
paper by \citet{zennaro19}, to which we refer the interested reader.

In a nutshell, the photometry of every cluster has been derived by
analyzing the images with the suite \textsc{Kitchen Sync 2}
\citep[see][]{sabbi16,bellini17}, originally developed by J. Anderson for the
analysis of ACS images \citep{anderson08a}.  The suite is comprised of FORTRAN
routines which recursively analyze the pixel values on the images and apply a
local-peak finding algorithm. In each analyzed image, a selection of relatively
bright stars is used to model a grid of spatially-variable point spread
functions (PSF), which represent the best-fit model for stars detected at
different locations on the chip.

The position and flux of stars with different luminosity are determined with two
different approaches. For bright stars, position and flux are independently
determined in each single exposure and then averaged, while for faint stars an
average position is computed from all the images where a star is detected, and
then the flux is determined at that position by using the best-fit PSF. 

Finally, stellar positions are corrected for geometrical distortion by using
the solution of \citet{bellini09b} and \citet{bellini11}. Instrumental
magnitudes are calibrated to the VEGAMAG system as in \citet{bedin05}. 

A selection of well-measured stars has been performed by using the parameters
of photometric quality produced by the reduction software as in
\citet{milone09}. The photometry of clusters affected by significant
differential reddening has been also corrected as in \citet{milone12c}.
Finally, the cluster membership has been established on the basis of relative
proper motions as done in \citet{milone18b}.

\subsection{Identification of AGB stars}\label{sec:selection}
A reliable identification of the AGB stars in each GC is a
critical step in our analysis. The procedure detailed below is based
on the observational features of AGB stars at different wavelengths.

The AGB phase begins at the onset of helium-shell burning. In low mass stars,
this phase is characterized by an electron-degenerate CO core whose mass
increases as the helium shell moves outwards. During the first evolutionary
stages, referred as early-AGB (EAGB), the \teff\ of the low mass AGB stars
steadily decreases while the luminosity increases, until they asymptotically
reach the typical values of RGB stars. From this moment on, EAGB stars evolve
by increasing their luminosity and decreasing \teff, with values slightly
higher than those of RGB stars, eventually becoming colder and brighter than
the brightest RGB stars near the AGB tip \citep[see e.g.][]{kamath12}.   

\begin{figure*}
\centering
\includegraphics[angle=270,width=\textwidth]{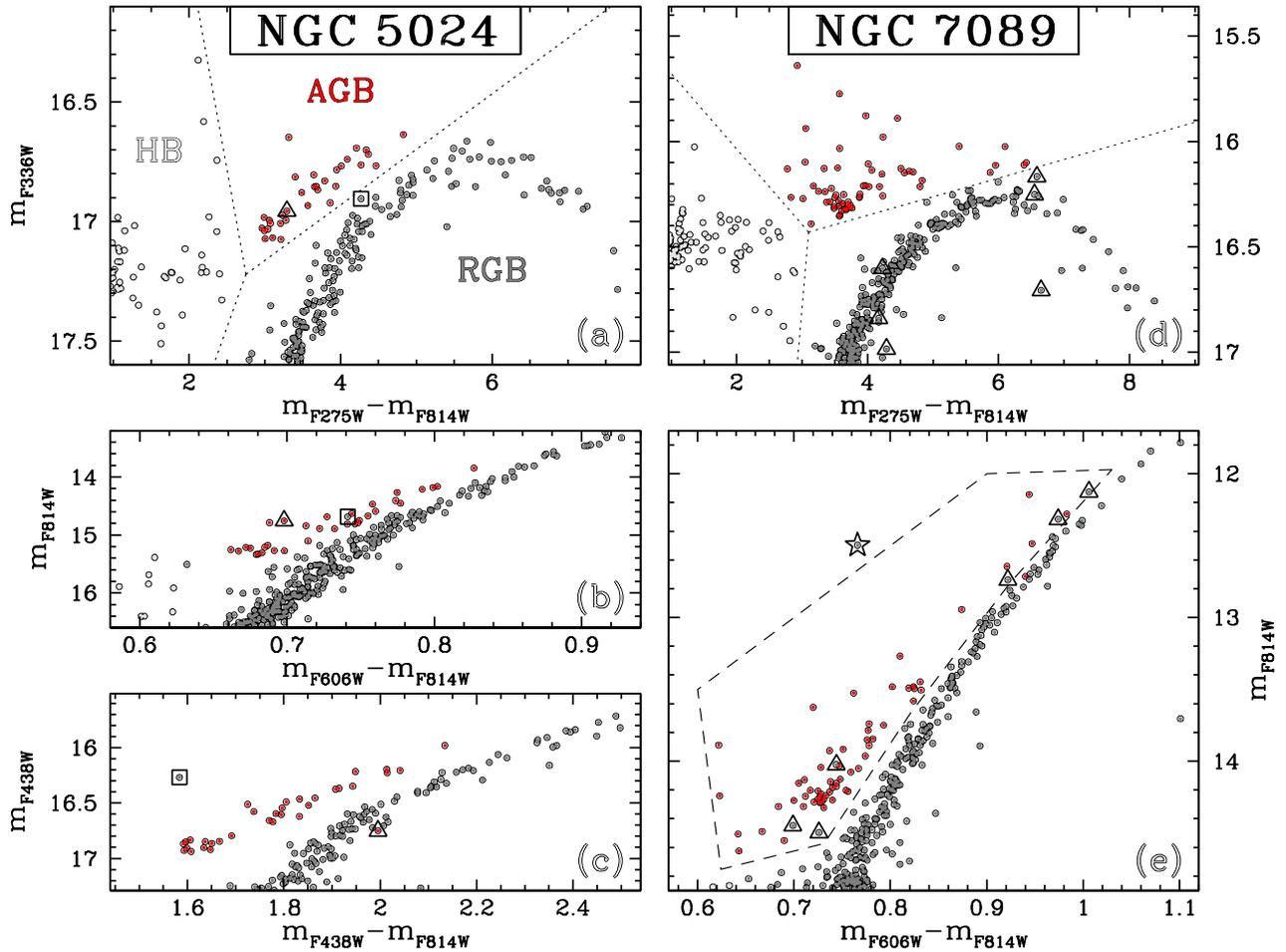} 
\caption{\textit{Panel a}: $m_{\rm F336W}$ vs. $(m_{\rm F275W}-m_{\rm F814W})$ CMD of the
	brightest stars of the Galactic cluster \ngc5024. The dotted lines
	divide the plot into three regions that include AGB, RGB, and HB stars,
	represented as red, dark-gray, and light-gray points, respectively.
	\textit{Panel b}: $m_{\rm F814W}$ vs. $(m_{\rm F606W}-m_{\rm F814W})$ CMD of the
	same stars plotted in panel (a). \textit{Panel c}: $m_{\rm F438W}$ vs.
	$(m_{\rm F438W}-m_{\rm F814W})$ CMD of the same stars in panel (a). Two AGB
	candidates with ambiguous membership attribution have been marked with
	geometric symbols. \textit{Panel d, e}: same
	as panel (a) but for the cluster \ngc7089. Stars marked with
	a starred symbol have location consistent with the AGB membership in the
	optical CMD of panel (e). A probable post-EAGB cluster star has
	been marked with an open circle. \label{fig:sel}} 
\end{figure*}

In optical CMDs, \teff\ variations between stars are visible as color
differences. At relatively high luminosities, random photometric errors are of
the order, or even larger, than the color differences corresponding to the
small temperature differences between AGB and RGB stars, thus making it
challenging to identify stars at different evolutionary stages. On the other
side, since the UV flux of EAGB stars is significantly higher than that of RGB
stars at similar luminosities, the two sequences appear well distinct in
far UV-optical CMDs \citep[][and references therein]{lagioia15a}. Such a behavior
is shown in Figure~\ref{fig:sel}, where we plotted the $m_{\rm F336W}$ vs.
$(m_{\rm F275W}-m_{\rm F814W})$ CMD (panel a) and the $m_{\rm F814W}$ vs.
$(m_{\rm F606W}-m_{\rm F814W})$ CMD (panel b) of the brightest stars of the Galactic GC
\ngc5024. In the first CMD, at $m_{\rm F336W} \lesssim 17.1$\,mag, we can
distinguish two sequences of stars, running almost parallel in the bottom-left
-- top right direction, separated by a color gap of less than 1\,mag: the
bluest and brightest sequence is populated by AGB stars, while the reddest by
RGB stars.  We can also recognize, at colors bluer than $\sim 2.5$\,mag, the
presence of a third group, mainly composed of HB stars. The distinct location
of these three stellar groups in the UV-optical CMD allows a straightforward
separation of the diagram into different regions, whose boundaries, marked by
the dotted lines in the plot, have been used to assign an evolutionary status
to the plotted stars.  For the sake of clarity, we decided to represent AGB,
RGB, and HB stars as red, dark-gray, and light-gray colored points,
respectively.  In passing, we notice that in this CMD, the increasing
F336W luminosity trend of RGB stars is reversed at $(m_{\rm F275W}-m_{\rm F814W})\gtrsim
6$\,mag because of the increasing importance of line blanketing by metals for
the coldest spectral type stars at UV wavelengths \citep[see][and references
therein]{buzzoni10}.

%%%%%%%%%%%%%%%%%%%%%%%%% FIGURE 1 %%%%%%%%%%%%%%%%%%%%%%%%%%%
%%%%%%%%%%%%%%%%%%%%%%%%%%%%%%%%%%%%%%%%%%%%%%%%%%%%%%%%%%%%%%

In the optical CMD of panel (b), we plotted with the same color code the
previously selected AGB, RGB and HB stars of \ngc5024. As mentioned before, the
determination of the AGB evolutionary membership in optical CMDs is thwarted by
the small (less than 0.05\,mag) color differences between AGB and RGB stars. We
notice, however, that in this CMD all the AGB stars selected in the UV-optical
CMD of panel (a) are bluer (hotter) than RGB stars at the same luminosity, thus
demonstrating that our method produces a reliable selection of
\textit{bona-fide} AGB stars.

Nonetheless, the selection procedure may also result in few ambiguous
attributions that are addressed by checking the location of equivocal cases in
CMDs obtained by using different band combinations. In the case of \ngc5024,
for instance, we can identify two of such stars. One star, marked by the black
triangle, that appears to be an AGB star in the CMD of panel (a) and (b), lies
almost beyond the red RGB boundary in the $m_{\rm F438W}$ vs.
$(m_{\rm F438W}-m_{\rm F814W})$ CMD displayed in panel (c). Another star, marked with
a square, lies at the border of the RGB region in the CMD of panel (a), and
appears as an AGB in the optical CMD of panel (b). However in panel (c) the
same star has a color too blue with respect to the rest of AGB stars.  This
comparison led us to exclude both the previous candidate AGBs from the \ngc5024
AGB sample.

The above procedure has been applied to identify the AGB candidates in every
cluster of our dataset except for a group of ten peculiar GCs. These clusters,
that in the extensive analysis of \citet{milone17} have been classified as
type\,II GCs, as opposed to the rest of clusters classified as type\,I, show a
significant internal variation of \feh\ and heavy elements \citep[][and references
therein]{marino19}. The photometric signature of this feature is revealed by
the presence of a secondary RGB sequence, redder than main one(s) in the 
$(m_{\rm F336W}-m_{\rm F814W})$ color \citep[see Figs. 10--18 in][]{milone17}. 
This group of RGB stars is usually referred to as anomalous population.

In the case of type\,II CMDs, our selection method may result in a spurious
bias that would exclude some AGB members belonging to the anomalous population.
To show this, we plot in panel (d) of Fig.~\ref{fig:sel}, the $m_{\rm F336W}$ vs.
$(m_{\rm F275W}-m_{\rm F814W})$ CMD of the type\,II GC \ngc7089 (M\,2). As for
\ngc5024, we defined three regions containing the majority of cluster AGB, RGB,
and HB stars, plotted with the previously adopted color code.  In the RGB
region, we can see a sparse group of stars fainter than the bulk of other RGBs,
composed by the cluster anomalous population.  However, in the $m_{\rm F814W}$ vs.
$(m_{\rm F606W}-m_{\rm F814W})$ CMD displayed in panel (e), the two RGB sequences are
indistinguishable. As a consequence, also the anomalous AGB stars will attain
the same colors of the other cluster AGB members. For this reason, we
empirically defined a region, indicated by the dashed line, that includes all
the stars with location compatible with the AGB phase. We observe that six
stars classified as RGB members in the UV-optical CMD, and marked with an open
starred symbol, fall in this region. We see that three of them lie along the
main RGB sequence in the UV-optical CMD of panel (d), two have F336W magnitudes
typical of the anomalous RGB sequence, while the remaining one lies at the
border of the RGB region.  Their location in both the CMDs indicate that they
are probable candidates of the AGB anomalous population and, therefore, they
have been merged to the cluster AGB sample. 

Interestingly, in the optical CMD we observe a star, that we marked with an
open circle, having two peculiar features: i) it is $\sim 0.2$\,mag bluer than AGB
stars at the same luminosity; ii) it has $m_{\rm F336W} \approx 14.6$, thus lying
beyond the displayed limits of the UV-optical CMD of panel (d). The bluer
optical color and the strong UV flux suggest that this star is a post-EAGB
candidate, namely a star with very small hydrogen-rich envelope, that fails to
reach the thermal-pulse AGB phase, leaving off the asymptotic branch towards
higher \teff\ \citep{greggio90}. We notice that the presence of a small mass
envelope is also connected with an higher mass loss during the RGB evolution,
which is in turn linked to higher helium content in stars \citep{tailo19a}.
\citet{milone18b} found that the maximum internal helium variation,
$\mathrm{\delta Y_{max}}$, of \ngc7089 is as high as $0.052$, therefore
corroborating the hypothesis on the peculiar nature of this star.

The method described above has been used for the selection of the sample of AGB
stars in all the others type\,II clusters, namely \ngc362, \ngc1261, \ngc1851,
\ngc5139 (\ome), \ngc5286, \ngc6715 (M\,54), \ngc6388, and \ngc6934.  

The total number of AGB stars, $\mathrm{N_{AGB}}$, identified in each cluster,
has been reported in Table~\ref{tab1} together with the total number of RGB
stars, $\mathrm{N_{RGB}}$, brighter than the faintest AGB candidate in F814W
band. We notice that since no AGB stars have been found in \ngc5053 and
\ngc6121, the corresponding $\mathrm{N_{RGB}}$ entries are
empty. In all the other 56 GCs, $\mathrm{N_{AGB}}$ ranges from 1
(\ngc6397) to 161 (\ngc6441). Among them, 21 GCs have $\mathrm{N_{AGB}} < 10$.
Since statistical significance is an important factor in the determination of
the photometric properties of AGB stars, we empirically assumed
$\mathrm{N_{AGB}} = 10$ as a lower limit for the inclusion of a GC in the
following analysis.

It is important here to stress the fact that the procedure for the selection of
the AGB candidates in type\,II GCs can results in a not complete definition of
the sample of their anomalous population. A better understanding of this
phenomenon requires, however, a deeper evaluation of the fraction of stars left
out by the adopted selection method, as done in the following section.

\subsection{Completeness of the AGB identifications in type\,II clusters}
In the previous section we have seen that in the $m_{\rm F336W}$ vs.
$(m_{\rm F275W}-m_{\rm F814W})$ CMD, the AGB sequence of the anomalous population on
type\,II GCs intersects the RGB sequence. This implies that part of the
anomalous AGB population can be left out by our selection procedure. To prevent
this issue, we have also seen that we can take advantage of the
$(m_{\rm F606W}-m_{\rm F814W})$ color, where the AGB sequences of all the cluster
stellar populations occupy the same region. Conversely, optical colors are far
from ideal for selecting AGB and RGB stars because the separation of the two
evolutionary sequences is comparable with the typical photometric error at
their characteristic luminosities. This implies that optical CMDs will only
provide a partial solution to the aforementioned issue.

It is therefore necessary to evaluate the maximum fraction of anomalous AGB
stars missed by our selection procedure. To this purpose, we performed a series
of simulations where we constructed artificial CMDs composed by stellar
populations with different iron abundance. To build the artificial CMDs, we
took advantage of appropriate evolutionary models from the
MESA~\footnote{\url{http://waps.cfa.harvard.edu/MIST/}} database
\citep{choi16}. In particular, we used isochrones of 13.5, 12.5, and 11.5 Gyr
at different \feh\ values, ranging from $-2.10$ to $-0.5$\,dex in steps of
0.15\,dex.  For each grid point, we simulated two artificial CMDs with
$\Delta$\feh$ = 0.15$ and 0.30\,dex.  The age and metallicity interval roughly
covers the entire range spanned by the studied Type\,II GCs, while the observed
average difference in \feh\ content between metal-rich and metal-poor stars in
these clusters is smaller than $\sim 0.35$\,dex \citep[][and references
therein]{marino19}.

\begin{figure*}
\centering
\includegraphics[angle=270,width=\textwidth,clip,trim={1.cm 0.3cm 9cm 2cm}]{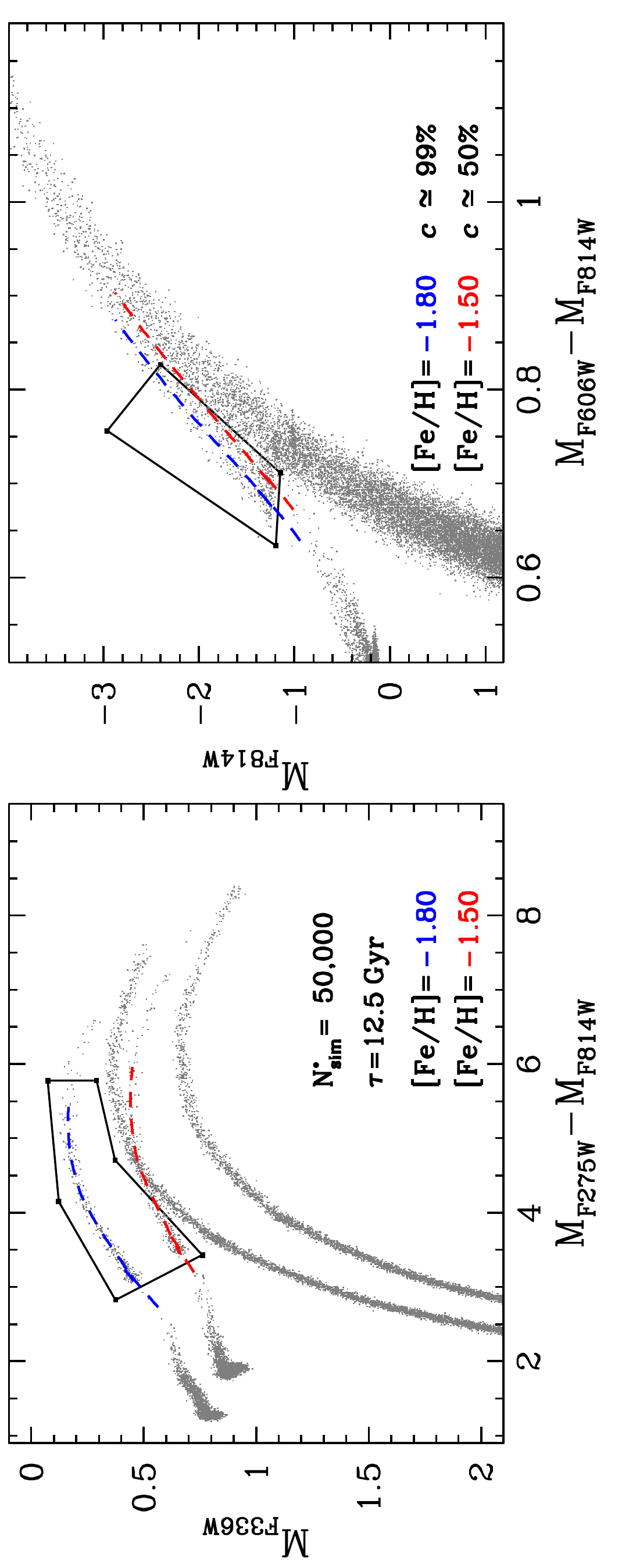}
\caption{Procedure for the estimate of the completeness of the AGB
	selection procedure in type\,II GCs. The left and right panel
	display, respectively, the $M_{\rm F336W}$ vs $M_{\rm F275W}-M_{\rm
	F814W}$ and the $M_{\rm F814W}$ vs. $M_{\rm F606W}-M_{\rm F814W}$ CMD
	of two 12.5\,Gyr-old synthetic stellar populations with \feh$=-1.80$
	and \feh$=-1.50$, each one composed by 50,000 stars. In each
	simulated population, the region including EAGB stars is represented by
	the corresponding isochrone (dashed line) whose metallicity has been
	indicated in the legend. In both the diagrams, the black boxes
	delimit the regions including the probable AGB members. Since
	part of the metal-rich AGB population mixes with the RGB of the
	metal-poor one in the first CMD, the completeness (\textit{c}) of the
	adopted selection procedure of the more metal-rich AGBs is
	correspondingly lower, as reported in the right panel legend.
	\label{fig:sim}}
\end{figure*}

Figure~\ref{fig:sim} shows an example of our simulations obtained from
12.5\,Gyr isochrones with \feh$=-1.80$ and \feh$=-1.50$, to which we will refer
as metal-poor and metal-rich population, respectively. Each simulated
population is composed of $5 \cdot 10^6$ stars with a random flat distribution
of initial masses higher than that of the corresponding model's
turn-off. Artificial magnitudes have been obtained by interpolation
on the initial mass values of the isochrone. To each simulated magnitude has
been assigned a random error based on a Gaussian distribution with the typical
dispersion of our observations. The resulting $M_{\rm F336W}$ vs. $M_{\rm
F275W}-M_{\rm F814W}$ and $M_{\rm F814W}$ vs.  $M_{\rm F606W}-M_{\rm F814W}$
artificial CMDs are displayed in the left and right panel of the figure, where
for the sake of clarity we decided to plot only 50,000 stars in each simulated
population.  Moreover, we marked the interval corresponding to the location of
the EAGB phase with a dashed line. In particular we used the blue color for the
metal-poor population, and the red color for the metal-rich one.

The left diagram shows that the metal-rich AGB sequence intersects the
metal-poor RGB one, so that the most evolved metal-rich AGBs appear fainter
than the brightest metal poor RGBs, in F336W band.  This fact clearly indicates
that the adopted AGB selection method can leave out part of the bright
metal-rich AGB population. In order to estimate the fraction of
missed stars, we performed the following steps: i) we tagged as input
AGB members the simulated stars with initial mass falling in the interval
defined by the upper and lower limit of the model EAGB range; ii) as done for
the observations, we defined by eye a region delimiting the probable AGB
candidates, as indicated by the black box in both the diagrams; iii) we flagged
as AGB members the stars falling in either box-delimited region; iv) we
compared the number of recovered and input AGB stars to estimate the fraction
of lost AGB stars.  By applying the above procedure for different combinations
of age and metallicity, we found that the maximum fraction of missed
AGB stars of the metal-rich stellar population is $\sim 1$\% when
$\Delta$\feh$=0.15$ and $\sim 50$\% when $\Delta$\feh$=0.30$,
corresponding respectively to a completeness of $\sim 99$\% and
$\sim 50$\%, as reported in the right panel. The missed
stars are mostly composed by AGB stars at later evolutionary stages.
Our simulations also show that a small fraction ($\sim 1$\%)
of metal-poor stars is left out by our selection: this derives from
the non-perfect location and extension of the AGB-delimiting region. We can
therefore consider 1\% as the random error associated to our
completeness estimates.

The most significant variations of completeness have been observed for
the most metal-poor and metal-rich bins of the analyzed metallicity range, and
for ${\rm \Delta\feh = 0.30}$. In the first case the completeness
of the metal-rich population is $\sim 30$\%, while it reduces to
nearly zero for the most metal-rich bin, where the metal-rich AGB is all
fainter and redder than the metal-poor RGB. We emphasize however that none of
the type\,II GCs in our GC database has metallicity lower than $\sim
-1.80$\,dex, and that for the most metal-rich Type\,II studied GC,
namely \ngc6388 \citep[${\rm \feh=-0.55}$, ][2010 update]{harris96a}, the
internal iron variation is negligible \citep[see][]{carretta18}.
Finally, we notice that similar results have been obtained by using
models of 13.5 and 11.5\,Gyr.

Our analysis indicates that the maximum fraction of anomalous AGB stars
missed by our selection procedure is negligible when the metallicity
difference between the anomalous and other GC stars is smaller than $\sim
0.15$\,dex. This fraction rapidly increases, becoming significant at
metallicity differences larger than $\sim 0.30$\,dex. Moreover, in the interval
considered in our simulations, cluster age has a negligible effect on the
results.

%%%%%%%%%%%%%%%%%%%%%%%%% FIGURE 2 %%%%%%%%%%%%%%%%%%%%%%%%%%%
%%%%%%%%%%%%%%%%%%%%%%%%%%%%%%%%%%%%%%%%%%%%%%%%%%%%%%%%%%%%%%%

\subsection{AGB candidates of the anomalous population in type\,II clusters}\label{sec:anomAGB}
In this section we seek AGB candidates of the anomalous population in type\,II
clusters \citep{milone17}. To do this we take advantage of their $m_{\rm
F336W}$ vs $m_{\rm F336W}-m_{\rm F814W}$ CMD. Indeed, as shown by
\citet{milone17}, this diagram represents an efficient tool to separate
populations with different metallicity in GCs.

%%%%%%%%%%%%%%%%%%%%%%%%% FIGURE 3 %%%%%%%%%%%%%%%%%%%%%%%%%%%
\begin{figure*}
\centering
\includegraphics[width=\textwidth]{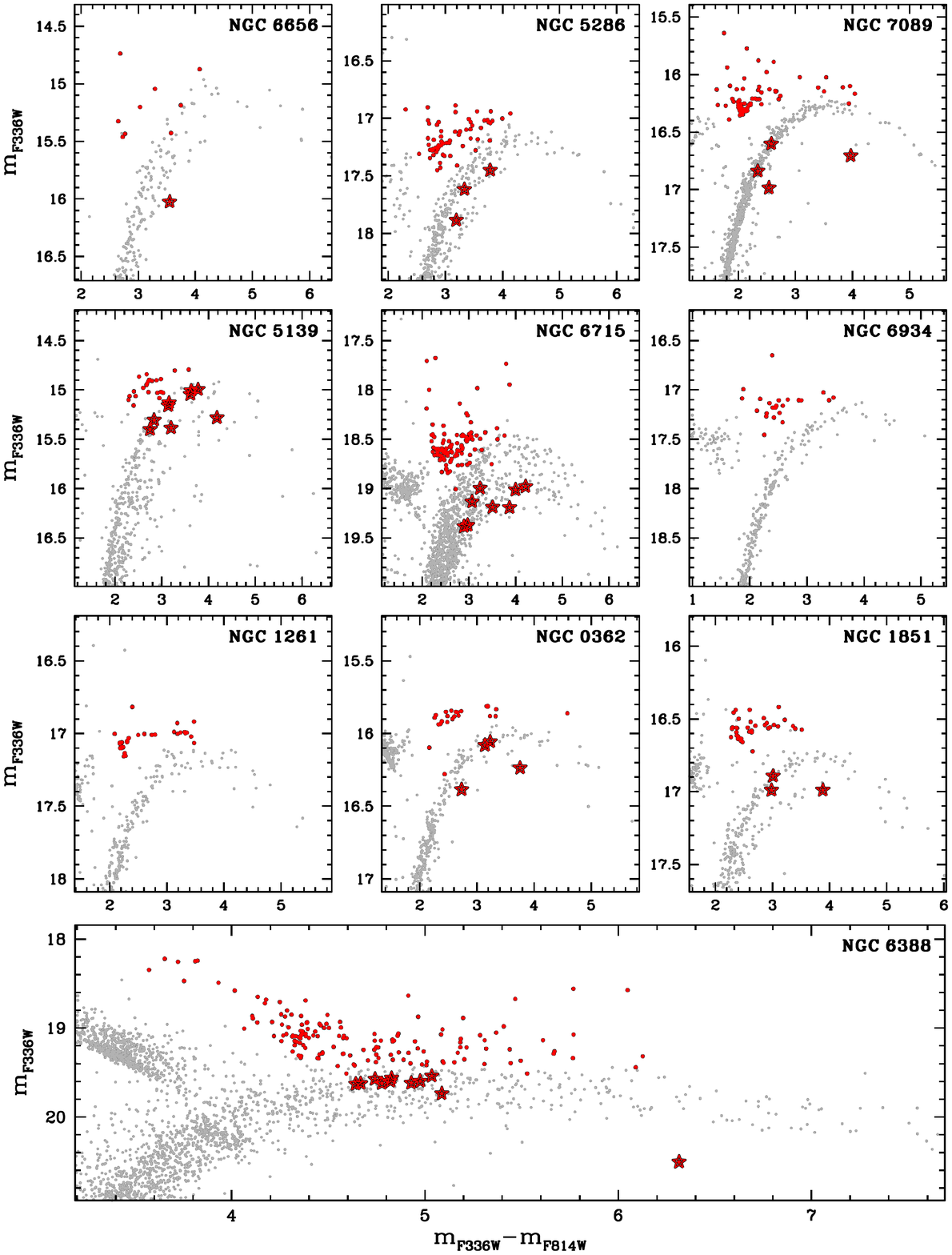}
\caption{$m_{\rm F336W}$ vs. $m_{\rm F336W}-m_{\rm F814W}$ CMDs of the ten type\,II
        clusters included in our database. In each GC, the AGB candidates are
        represented as red points, while the AGB candidates of the anomalous
        population have been marked with a starred symbol.\label{fig:TypeIIcmds}}
\end{figure*}
%%%%%%%%%%%%%%%%%%%%%%%%%%%%%%%%%%%%%%%%%%%%%%%%%%%%%%%%%%%%%%%

Figure~\ref{fig:TypeIIcmds} shows the $m_{\rm F336W}$ vs $m_{\rm F336W}-m_{\rm F814W}$ CMDs of
the ten Type\,II GCs included in our database, arranged from the less
to the most metal-rich. In each CMD, we marked with red points the
AGB candidates found through the selection procedure described in
Sect.~\ref{sec:selection}.  We see that in all the CMDs the majority of AGB
members occupy the uppermost portion of the diagram, except for a group of few
stars that have F336W luminosities comparable with or lower than that of RGB
stars at the same colors. As seen in the previous section, this indicates that
they are likely to be members of the AGB anomalous population. However, since
our AGB selection procedure is mainly based on the location of stars in the
$m_{\rm F336W}$ vs $m_{\rm F275W}-m_{\rm F814W}$ CMD, some of the AGB stars not belonging
to the anomalous population may attain colors similar to those of RGB stars in
the narrower color-baseline $m_{\rm F336W}$ vs $m_{\rm F336W}-m_{\rm F814W}$ CMD.
According to this consideration, we identified as the most probable members of
the anomalous AGB population, only those AGB stars fainter-redder than the main
RGB, while the AGBs lying at the blue border of the main RGB sequence have been
excluded. In the case of \ngc5139 and \ngc6388, since no clear separation is
visible between the different cluster RGB sequences, we decided to flag as
anomalous AGBs all the stars redder - fainter than the blue limit of the main
RGB sequence. Since the majority of anomalous AGB candidates in \ngc6388 attain
similar $m_{\rm F336W}-m_{\rm F814W}$ colors, we decided to use a wider panel in order
to stretch the horizontal scale of the plot.  In each panel, the selected
anomalous AGB candidates have been marked with an open starred symbol in each
CMD. We notice that in two clusters, namely \ngc1261 and \ngc6934, no anomalous
AGBs were found, while only one anomalous AGB candidate has been identified in
\ngc6656, which is one of the clusters with the highest fraction ($\sim 0.4$)
of anomalous population RGB stars among those analyzed by \citet{milone17}. On
the other side, the highest number of AGB candidates have been detected in the
three massive GCs \ngc5139 (9) \ngc6388 (11), and \ngc6715 (8). 

\begin{figure}	
\centering
\includegraphics[width=\columnwidth]{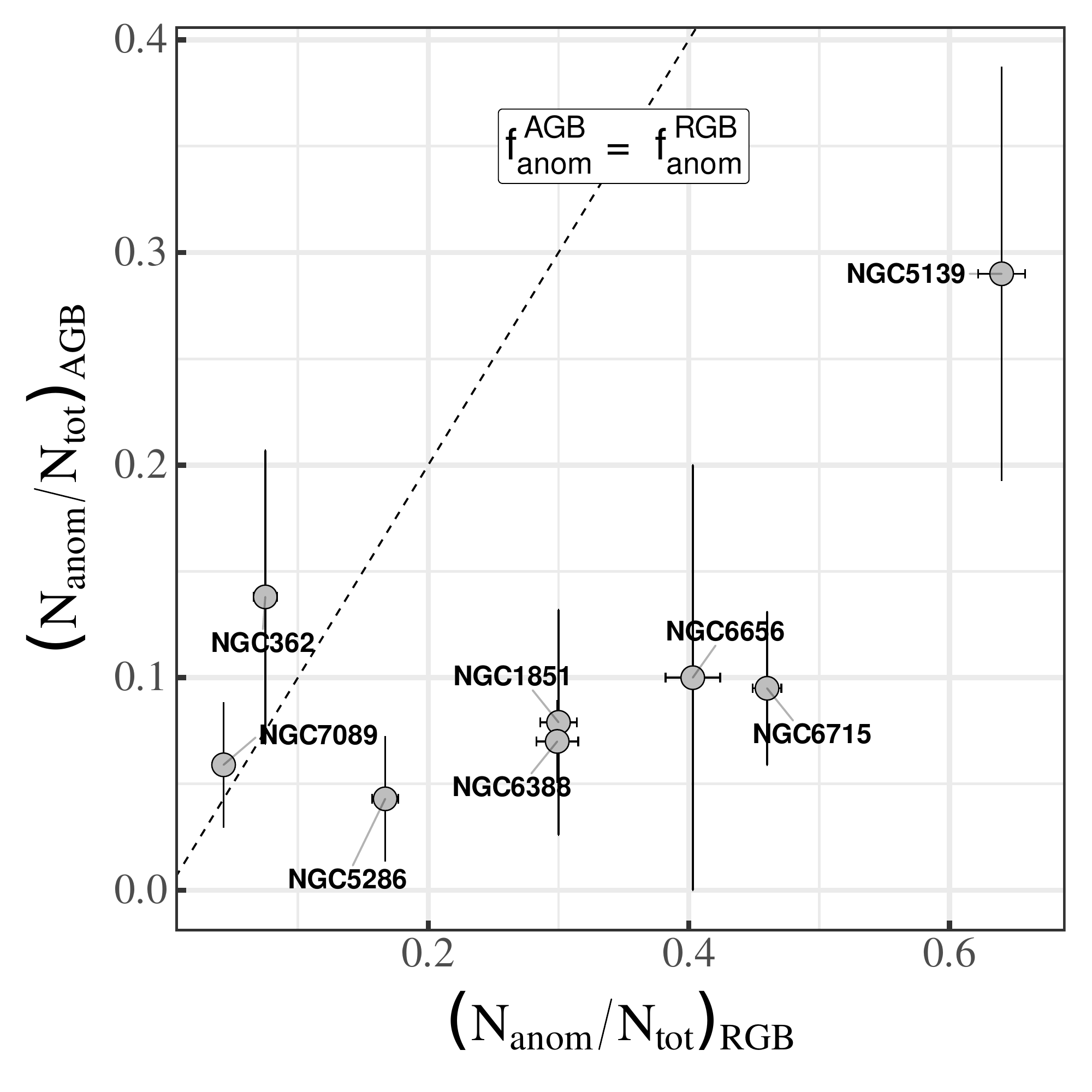} 
\caption{Fraction of anomalous AGB stars vs. fraction of anomalous RGB stars of
	eight type\,II clusters included in our database. The black dashed line
	represents the identity relation. \label{fig:f_anom}}
\end{figure}

Figure ~\ref{fig:f_anom} shows the relation between the fraction of anomalous
candidate AGBs, $\mathrm{(N_{anom}/N_{tot})_{AGB}}$, as a function of the
fraction of anomalous RGB stars $\mathrm{(N_{anom}/N_{tot})_{RGB}}$ determined
by \citet{milone17}. For each cluster $i$, the error associated to the estimate
of the fraction of anomalous AGBs has been obtained by performing 50,000
simulations of $\mathrm{(N^i_{tot})_{AGB}}$ values randomly distributed in the
interval $[0,1]$. For each simulation, we evaluated the fraction of simulated
values smaller than the $i$-th $\mathrm{(N_{anom}/N_{tot})}_{AGB}$ value.
Finally the 68th percentile of the resulting distribution has been taken as the
error associated to the anomalous AGB fraction. The black dashed line in the
plot marks the identity relation.

We see that in six GCs the fraction of anomalous RGBs is significantly higher
than that of the AGBs, thus suggesting that part of their anomalous population
does not reach the asymptotic branch. In \ngc362 and \ngc7089 the fraction of
anomalous AGBs is higher, but consistent within one sigma with that of RGB
stars. We also notice that these two clusters are among the Type\,II GCs with
the lowest fraction of anomalous RGB stars (respectively $\sim 8$\% and $\sim
4$\%; \citeauthor{milone17}, \citeyear{milone17}). As a consequence, small
random statistical fluctuations can significantly affect the estimate of their
fraction of anomalous AGBs.

%%%%%%%%%%%%%%%%%%%%%%%%% FIGURE 4 %%%%%%%%%%%%%%%%%%%%%%%%%%%
%%%%%%%%%%%%%%%%%%%%%%%%%%%%%%%%%%%%%%%%%%%%%%%%%%%%%%%%%%%%%%%

\subsection{Determination of the AGB and RGB width}\label{sec:width}
The photometric footprint of the presence of MPs in GCs is represented by the
color spread of stars, observed in CMDs where specific color combinations,
sensitive to the content variation of light-elements and helium, are used. The
two main choices adopted so far are the index $C_{\rm F275W,F336W,F438W}$ and
$C_{\rm F336W,F438W,F814W}$: the first mostly traces chemical differences in O, N,
and C, while the second in N, C and helium.

\begin{figure*}
\centering
\includegraphics[angle=270,width=\textwidth]{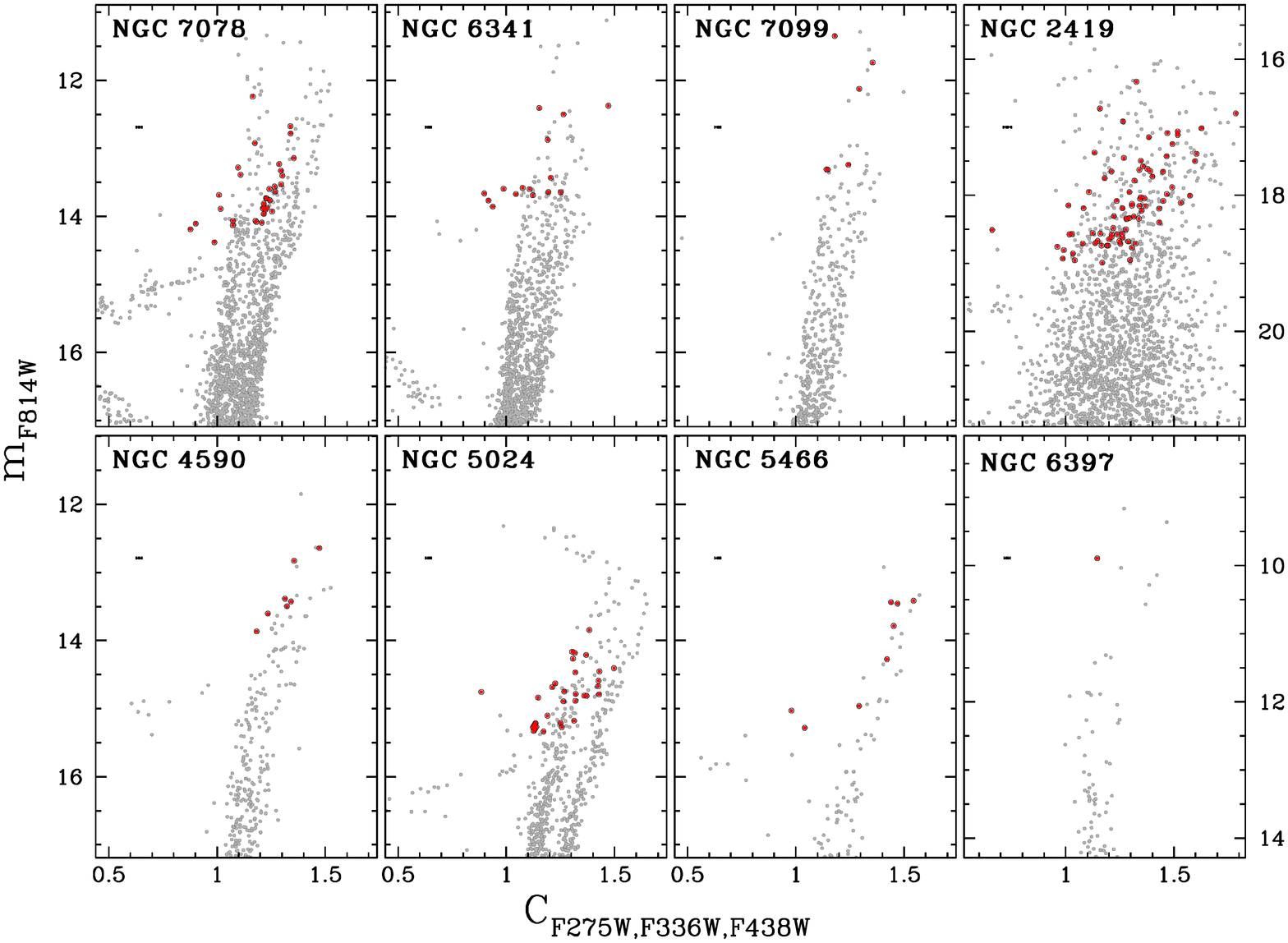} 
\caption{$m_{\rm F814W}$ vs. $C_{\rm F275W,F336W,F438W}$ of \ngc7078, \ngc6341,
	\ngc7099, \ngc2419, \ngc4590, \ngc5024, \ngc5466, and \ngc6397. AGB
	stars are represented as red points and their average error bar 
	is shown on the upper-left side of each panel.
	\label{fig:cmds1}}
\end{figure*}

\begin{figure*}
\centering
\includegraphics[angle=270,width=\textwidth]{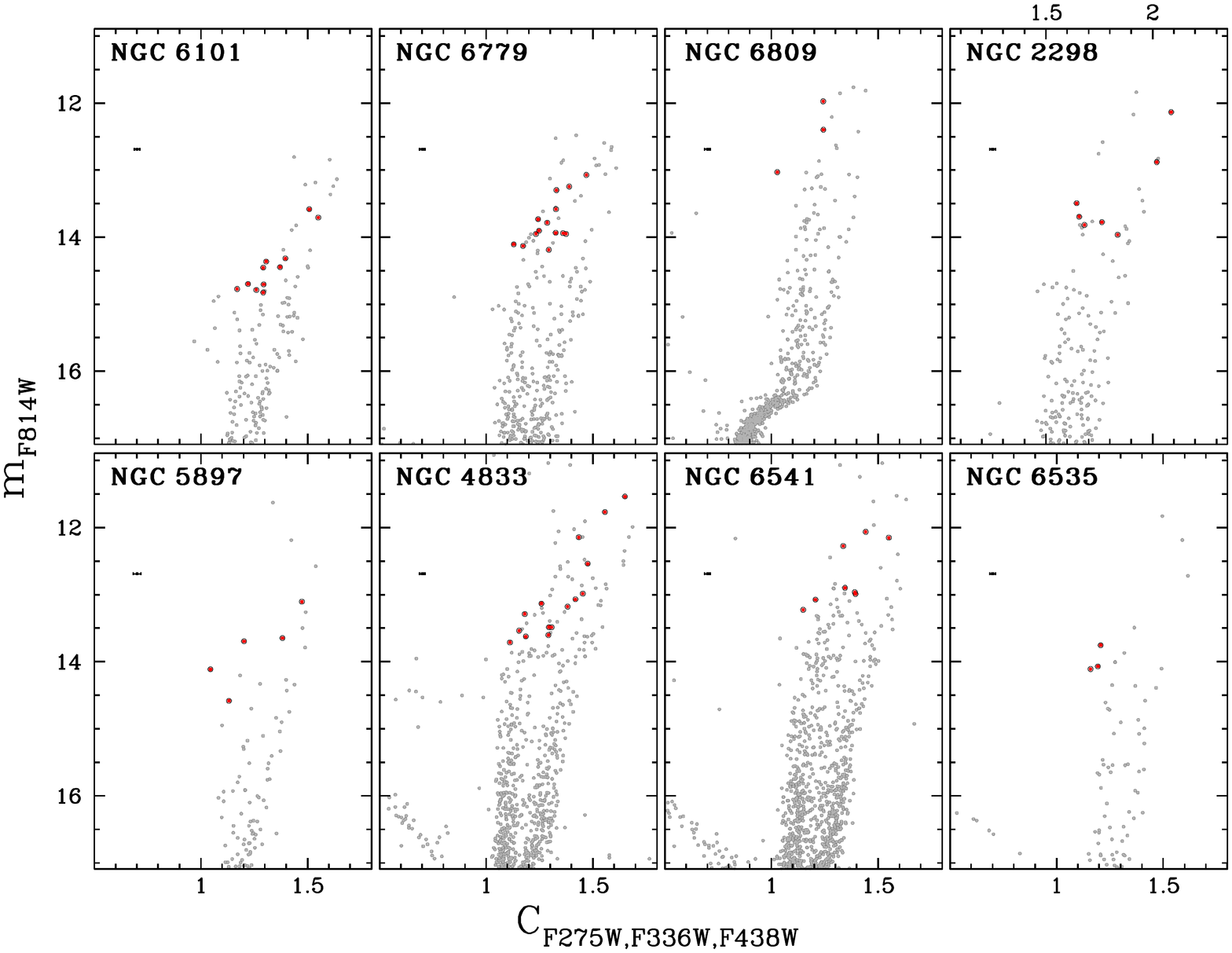} 
\caption{Same as Fig.~\ref{fig:cmds1} but for \ngc6101, \ngc6779, \ngc6809,
	\ngc2298, \ngc5897, \ngc4833, \ngc6541,  and \ngc6535.\label{fig:cmds2}}
\end{figure*}

\begin{figure*}
\centering
\includegraphics[angle=270,width=\textwidth]{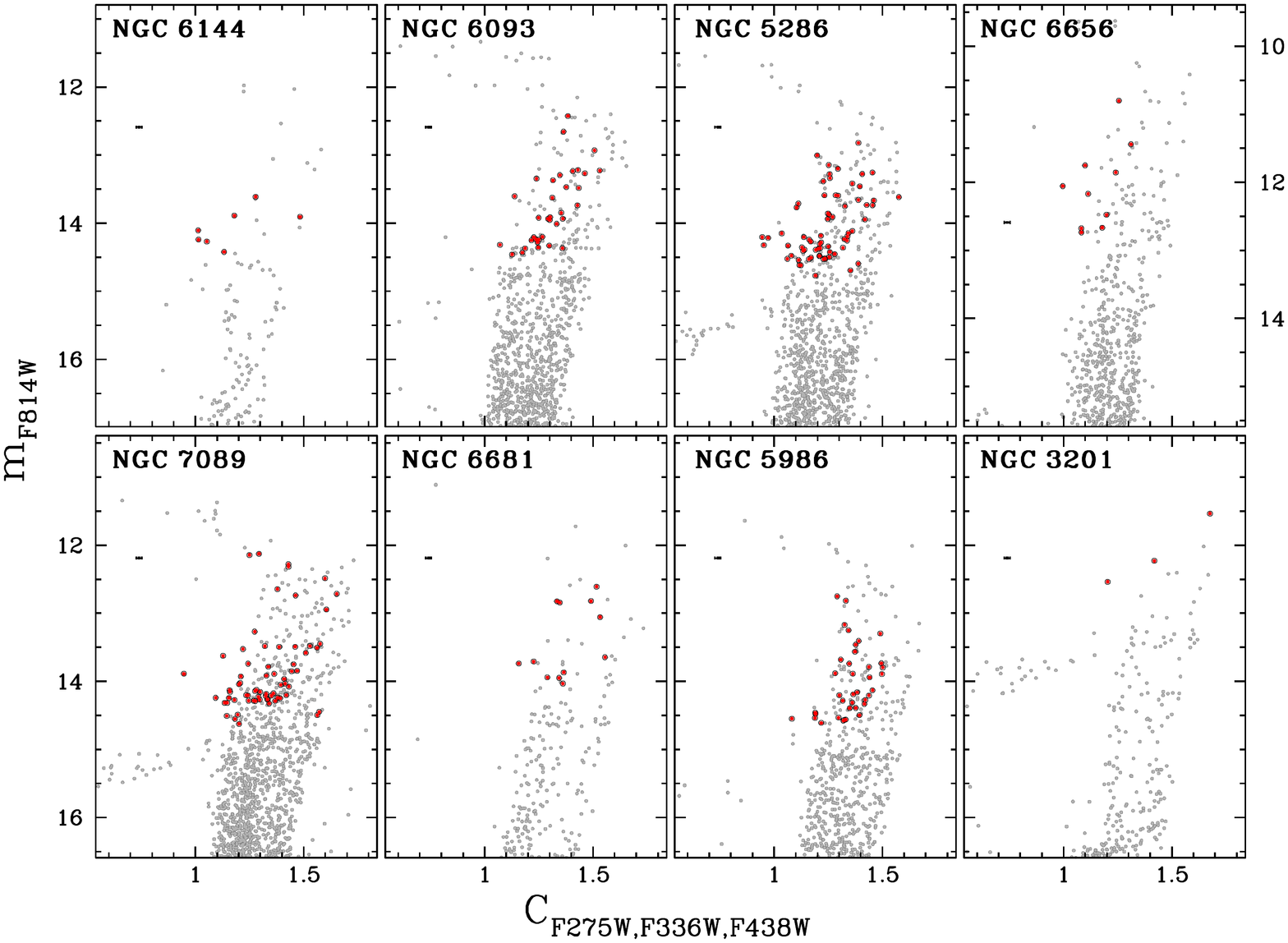} 
\caption{Same as Fig.~\ref{fig:cmds1} but for \ngc6144, \ngc6093, \ngc6656,
	\ngc5286, \ngc7089, \ngc6681, \ngc5986, and \ngc3201.\label{fig:cmds3}}
\end{figure*}

\begin{figure*}
\centering
\includegraphics[angle=270,width=\textwidth]{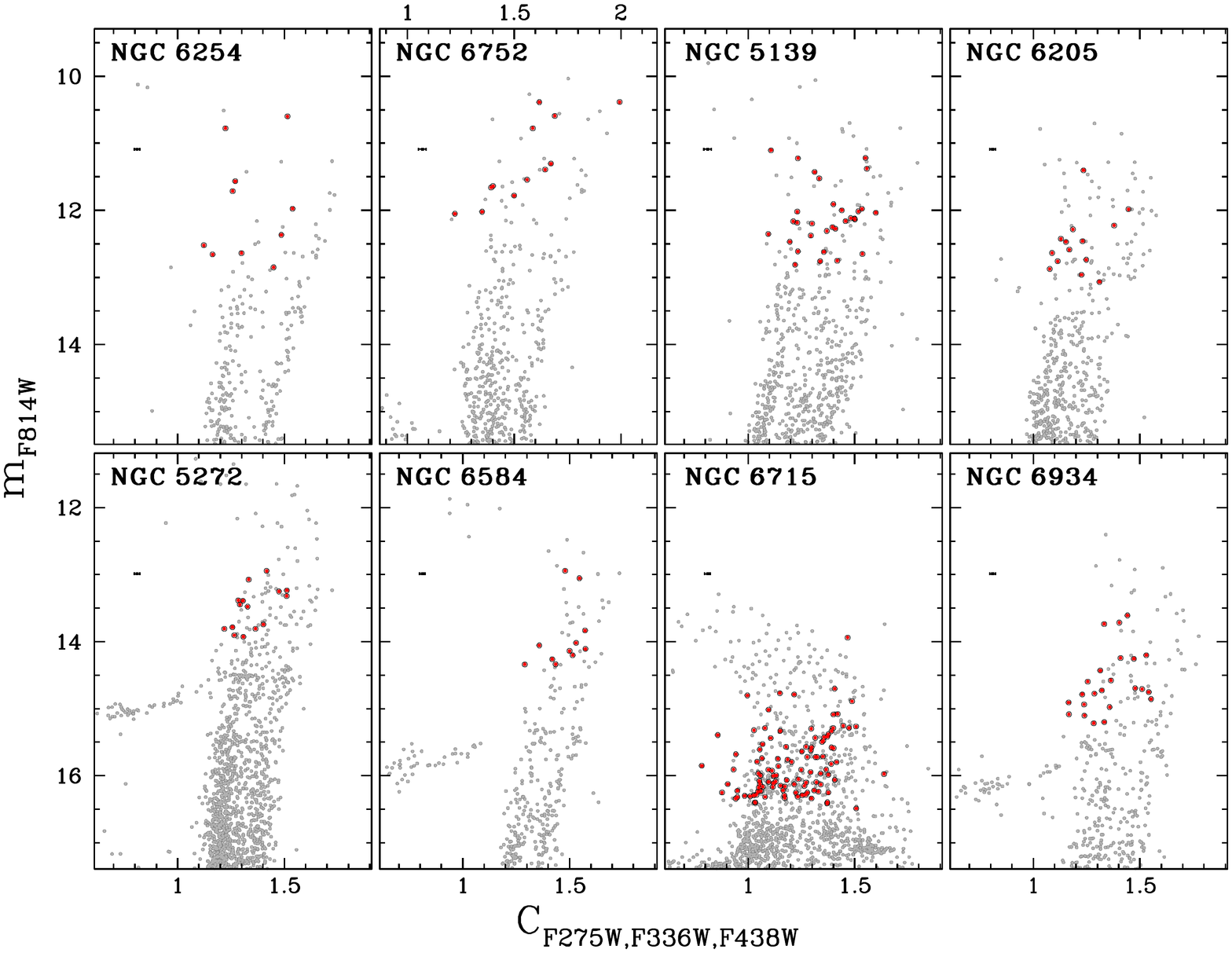} 
\caption{Same as Fig.~\ref{fig:cmds1} but for \ngc6254, \ngc6752 \ngc5139,
	\ngc6205, \ngc5272, \ngc6584, \ngc6715, and \ngc6934.\label{fig:cmds4}}
\end{figure*}

\begin{figure*}
\centering
\includegraphics[angle=270,width=\textwidth]{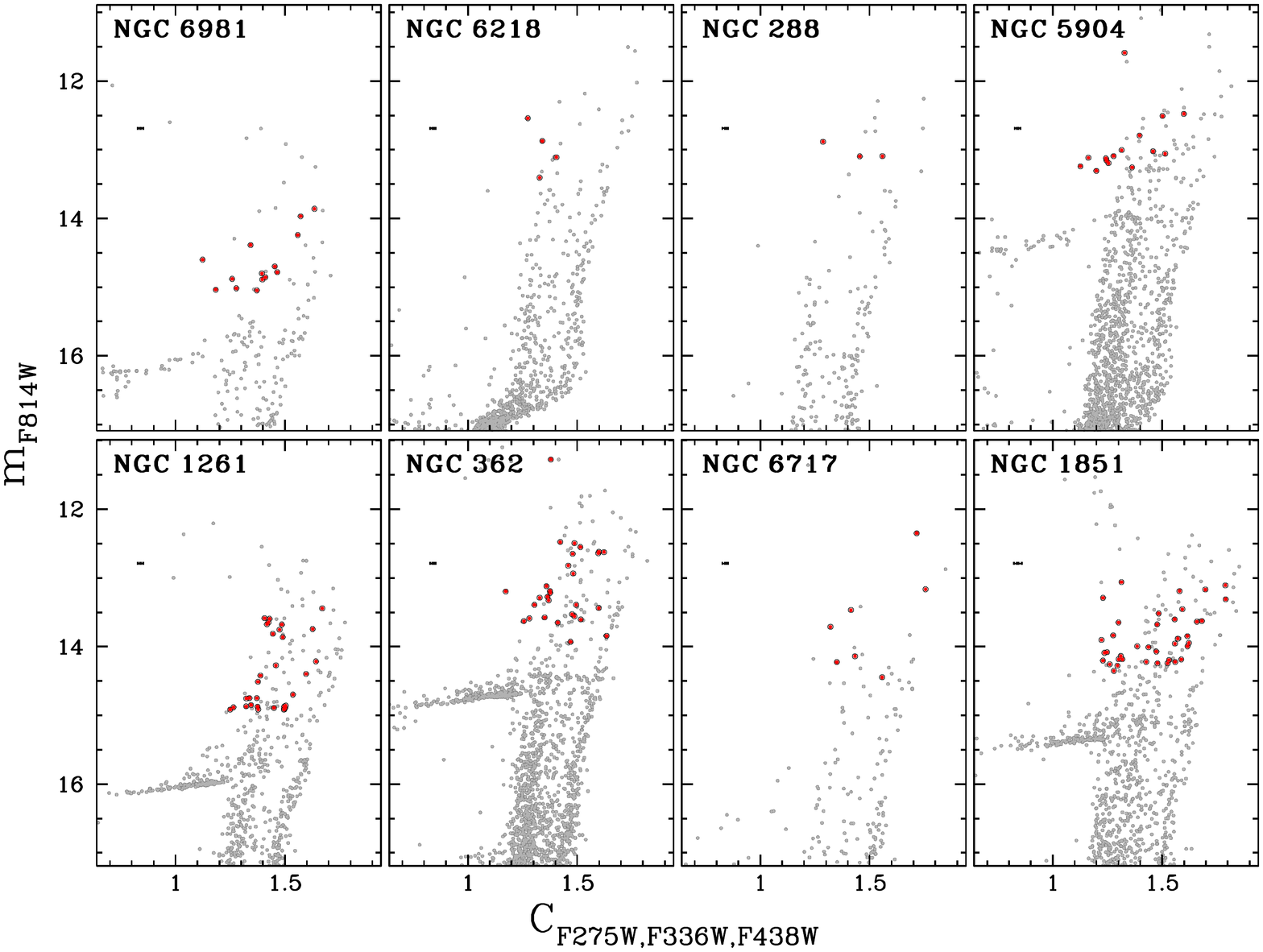} 
\caption{Same as Fig.~\ref{fig:cmds1} but for \ngc6981, \ngc6218, \ngc288,
	\ngc5904, \ngc1261, \ngc362, \ngc6717, and \ngc1851.\label{fig:cmds5}}
\end{figure*}

\begin{figure*}
\centering
\includegraphics[angle=270,width=\textwidth]{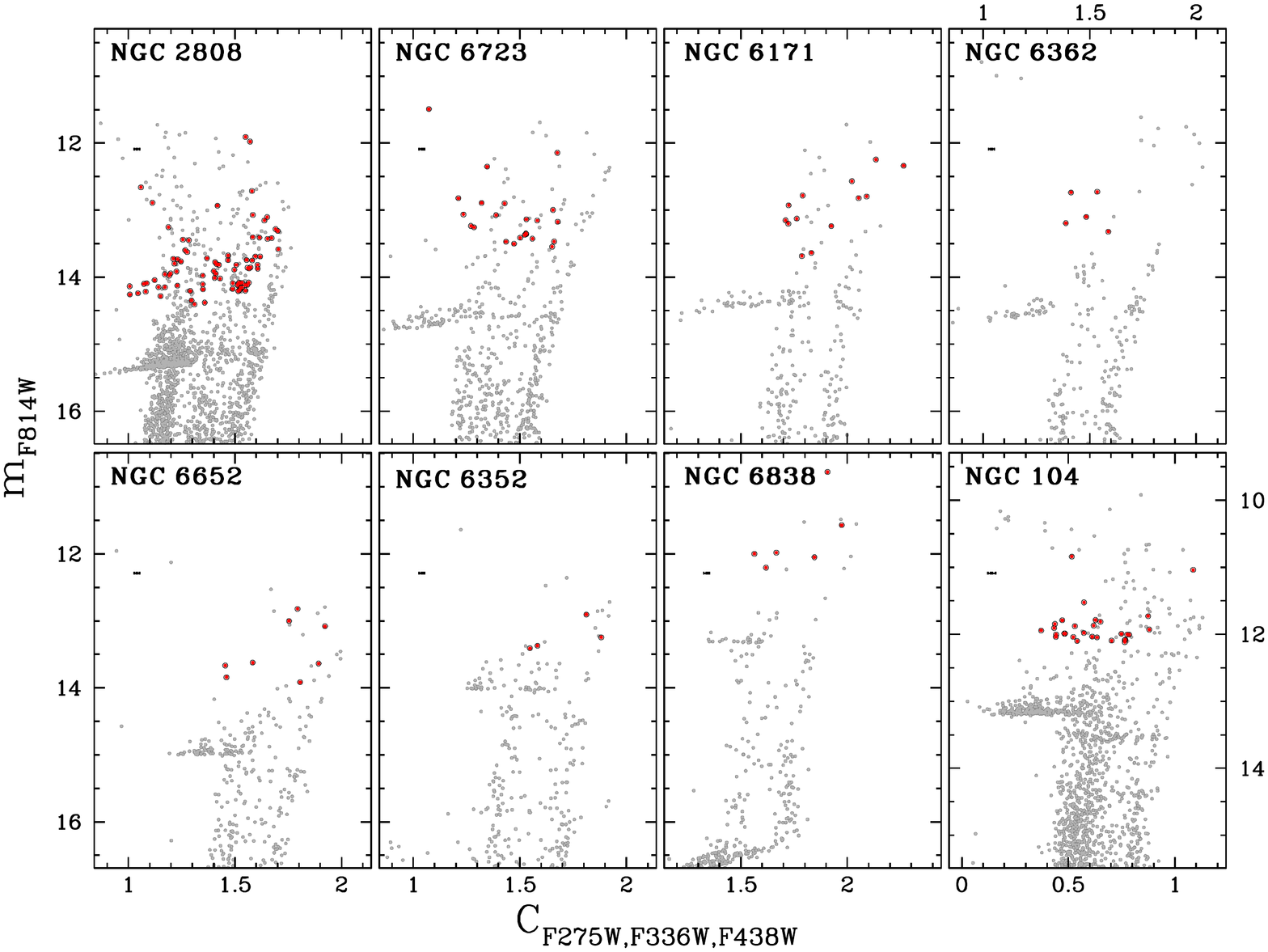} 
\caption{Same as Fig.~\ref{fig:cmds1} but for \ngc2808, \ngc6723, \ngc6171,
	\ngc6362, \ngc6652, \ngc6352, \ngc6838, and \ngc104.\label{fig:cmds6}}
\end{figure*}

\begin{figure*}
\centering
\includegraphics[angle=270,width=\textwidth]{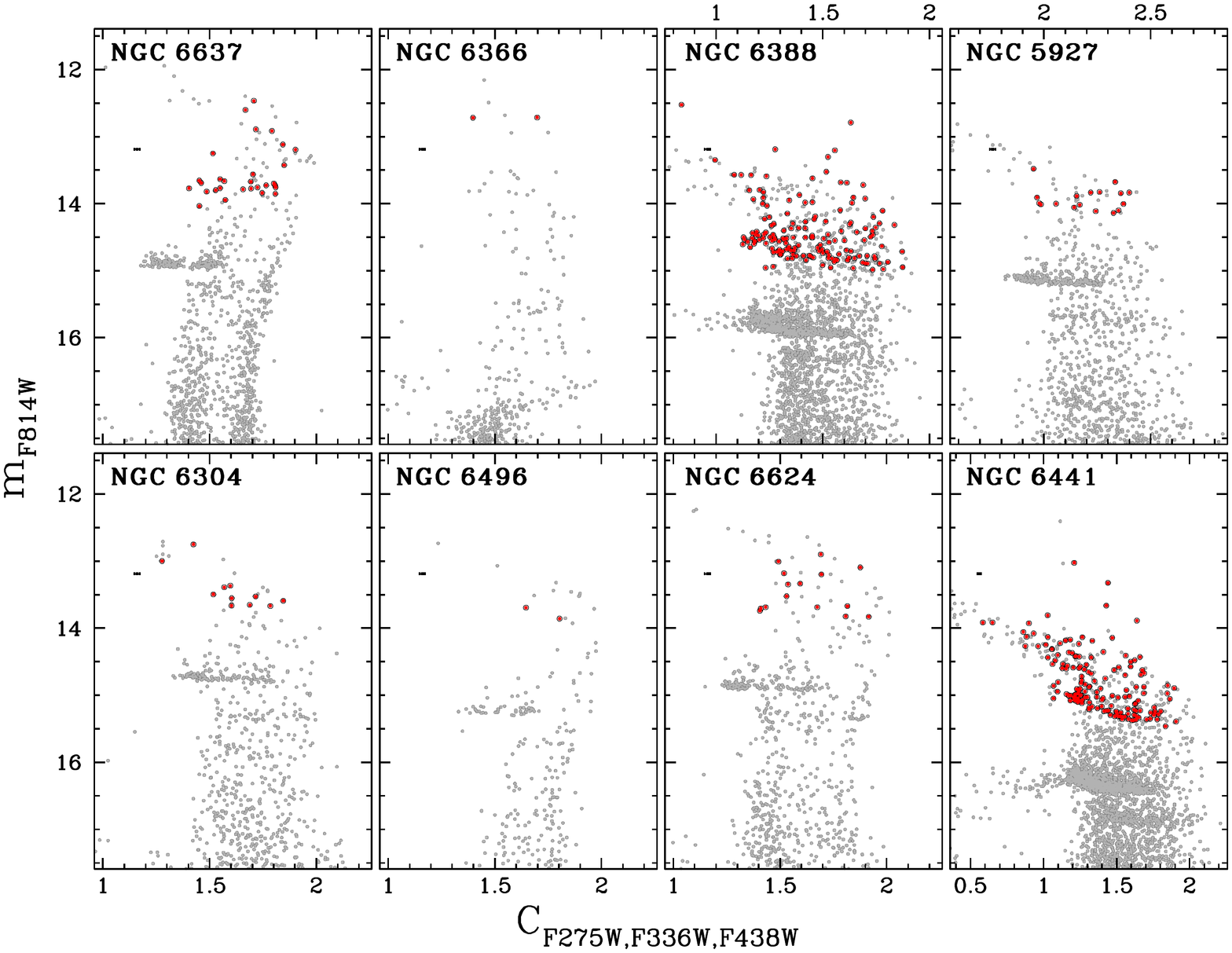} 
\caption{Same as Fig.~\ref{fig:cmds1} but for \ngc6637, \ngc6366, \ngc6388,
	\ngc5927, \ngc6304, \ngc6496, \ngc6624, and \ngc6441.\label{fig:cmds7}}
\end{figure*}

The $m_{\rm F814W}$ vs. $C_{\rm F275W,F336W,F438W}$ pseudo-CMDs of the brightest stars
in all the analyzed GCs, sorted in ascending order of metallicity, are
displayed in Figures~\ref{fig:cmds1}, \ref{fig:cmds2}, \ref{fig:cmds3}, \ref{fig:cmds4},
\ref{fig:cmds5}, \ref{fig:cmds6} and \ref{fig:cmds7}. For the sake of convenience we
decided to not show also the $m_{\rm F814W}$ vs. $C_{\rm F336W,F438W,F814W}$
pseudo-CMDs of our targets. In each CMD, the gray points
represent the cluster RGB and HB stars, while the red points the AGB
candidates. The black error bars on the left side of each panel indicate the
typical photometric errors of the AGB stars. 
%(To be deleted?): derived from the analysis of
%\ngc2808, which has been used as a template cluster. Indeed, since the majority
%of the observed clusters have a relatively small population of AGB stars, it
%would be hard to obtain a reliable estimate of the corresponding photometric
%errors. To compute the photometric error of the 83 AGB stars of \ngc2808 we
%used the procedure reported in \citet{lagioia19a}. In a nutshell, for each band
%we derived the corresponding magnitude dispersion through the relation
%$\mathrm{\sigma^i_X = rms^i_X/\sqrt{N_X-1}}$, with X = F275W, F336W, F438W,
%F814W and N = number of magnitude measurement of the $i$-th star. The
%sigma-clipped median value of the $\mathrm{\sigma^i_X}$ observations has then
%been used to simulate a normal distribution of $10^5$ artificial errors that
%takes into account the contribution of the differential reddening correction
%and zero-point variations. The 68.27th percentile of the resulting distribution
%has been taken as our error estimate. We found that the typical error in F814W
%band for the AGB stars is 0.008\,mag while that in the pseudo-color
%$C_{\rm F275W,F336W,F438W}$ is 0.013\,mag. For the six GCs in our database for no
%differential reddening correction has been applied, the two previous value are,
%respectively, 0.009\,mag and 0.018\,mag.

A quick look at the CMDs shows that the AGB sequences are much wider than the
spread expected from observational errors alone in all the analyzed GCs, except
in eight clusters, namely \ngc7099, \ngc4590, \ngc5466, \ngc6397, \ngc6809,
\ngc6535, \ngc6218, and \ngc6496, that have a poorly populated sequence of AGB
stars with a spread comparable with the pseudo-color error. On the other hand,
the poorly populated AGB sequence of \ngc2298, \ngc5897, \ngc6541, \ngc6144,
\ngc3201, \ngc288, \ngc6717, \ngc6362, \ngc6352, \ngc6838, and \ngc6366
displays a clear spread. The pseudo-color broadening observed in the
CMDs clearly show that MPs are a common characteristic of the analyzed GCs.

%%%%%%%%%%%%%%%%%%%%%%%%% FIGURE 5 %%%%%%%%%%%%%%%%%%%%%%%%%%%
%%%%%%%%%%%%%%%%%%%%%%%%%%%%%%%%%%%%%%%%%%%%%%%%%%%%%%%%%%%%%%%

%%%%%%%%%%%%%%%%%%%%%%%%% FIGURE 6 %%%%%%%%%%%%%%%%%%%%%%%%%%%
%%%%%%%%%%%%%%%%%%%%%%%%%%%%%%%%%%%%%%%%%%%%%%%%%%%%%%%%%%%%%%%

%%%%%%%%%%%%%%%%%%%%%%%%% FIGURE 7 %%%%%%%%%%%%%%%%%%%%%%%%%%%
%%%%%%%%%%%%%%%%%%%%%%%%%%%%%%%%%%%%%%%%%%%%%%%%%%%%%%%%%%%%%%%

%%%%%%%%%%%%%%%%%%%%%%%%% FIGURE 8 %%%%%%%%%%%%%%%%%%%%%%%%%%%
%%%%%%%%%%%%%%%%%%%%%%%%%%%%%%%%%%%%%%%%%%%%%%%%%%%%%%%%%%%%%%%

%%%%%%%%%%%%%%%%%%%%%%%%% FIGURE 9 %%%%%%%%%%%%%%%%%%%%%%%%%%%
%%%%%%%%%%%%%%%%%%%%%%%%%%%%%%%%%%%%%%%%%%%%%%%%%%%%%%%%%%%%%%%

%%%%%%%%%%%%%%%%%%%%%%%%% FIGURE 10 %%%%%%%%%%%%%%%%%%%%%%%%%%%
%%%%%%%%%%%%%%%%%%%%%%%%%%%%%%%%%%%%%%%%%%%%%%%%%%%%%%%%%%%%%%%

%%%%%%%%%%%%%%%%%%%%%%%%% FIGURE 11 %%%%%%%%%%%%%%%%%%%%%%%%%%%
%%%%%%%%%%%%%%%%%%%%%%%%%%%%%%%%%%%%%%%%%%%%%%%%%%%%%%%%%%%%%%%

\begin{figure*}
\centering
\includegraphics[angle=270,width=\textwidth]{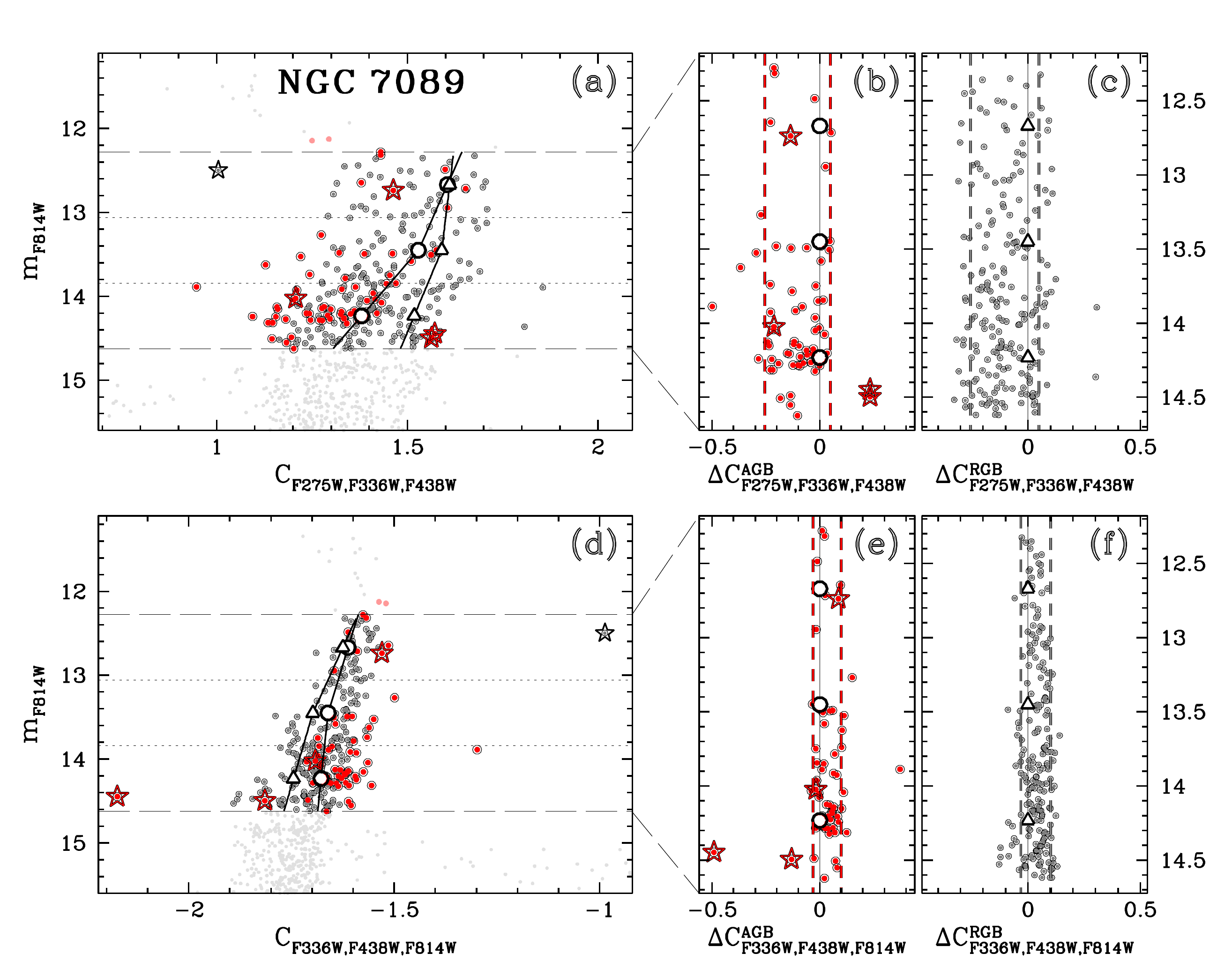} 
\caption{Determination of the width of AGB and RGB stars of \ngc7089.
	\textit{Panel (a)}: $m_{\rm F814W}$ vs. $C_{\rm F275W,F336W,F438W}$ CMD of the
	cluster AGB (red points) and RGB/HB (gray points) stars included in the
	selected F814W magnitude interval marked by the long-dashed horizontal
	lines. AGB stars brighter than than the selected luminosity threshold
	are represented as light-red points. Anomalous AGB candidates have been
	marked with a red starred symbol, while the only post-EAGB candidate
	star has been marked with a black starred symbols. \textit{Panel (b),
	(c)}: `verticalized' CMDs of the cluster AGB (panel b) and RGB (panel
	c) stars, where the horizontal displacement between the two dashed
	vertical lines represents the corresponding observed width.
	\textit{Panel (d), (e), (f)}: same as in panel (a), (b), (c) but for
	the pseudo-color $C_{\rm F336W,F438W,F814W}$.  \label{fig:wM15}}
\end{figure*}

A deeper view into the properties of MPs can be obtained through the analysis
of their color extension. Figure ~\ref{fig:wM15} illustrates the procedure for
the measurement of the width of the AGB and the upper RGB stars of \ngc7089.
Panel (a) shows the $m_{\rm F814W}$ vs. $C_{\rm F275W,F336W,F438W}$ pseudo-CMD
of the brightest cluster stars, where gray and red points represent
respectively RGB/HB members and AGB candidates.  We also highlighted the
anomalous AGB stars of the clusters with the usual red starred symbol. So we
did with the post-EAGB candidate, marked with a black starred symbol. Stars in
this diagram follow an almost vertical trend along all the displayed magnitude
interval until $\approx 12.2$\,mag, where they move towards bluer colors
because of the effect of line blanketing, as seen in the previous section. As
such, all the AGB/RGB stars brighter than this threshold have been excluded
from the following computation.  We therefore defined a magnitude interval,
delimited by the long-dashed horizontal lines, extended from the brightest to
the faintest cluster AGB star, not belonging to the cluster anomalous AGB
sample, which has been discarded in the following procedure.  We then divided
this luminosity range in three magnitude bins, delimited by the dotted
horizontal lines. For each bin \textit{i}, we measured the 80th percentile of
the pseudo-color distribution of the AGB and of the RGB star sample, indicated
respectively by open circles and triangles.  Then, we linearly interpolated
these points along the magnitude values attained by the AGB and by the RGB
stars included in the selected magnitude interval. The interpolation line was
then used to `verticalize' the CMD of the AGB stars (panel b) and of the RGB
stars (panel c). In these two diagrams, the abscissa of each star, ${\rm
\Delta}_{C\,{\rm F275W,F336W,F438W}}^{\rm AGB,\,obs}$, and ${\rm
\Delta}_{C\,{\rm F275W,F336W,F438W}}^{\rm RGB,\,obs}$, corresponds to the
difference between its pseudo-color and the pseudo-color of the interpolating
function at the same F814W magnitude, whose abscissa is now identically equal
to zero.  Finally, we computed the 10th and the 90th percentile of the ${\rm
\Delta}_{C\,{\rm F275W,F336W,F438W}}^{\rm AGB\,(RGB)}$ distribution, marked by
the red (gray) short-dashed vertical lines in panel b (c), and took the
absolute difference between these two values as the width of the cluster AGB
(RGB) stars, $W^{\rm AGB\,(RGB),\,obs}_{C\,{\rm F275W,F336W,F438W}}$.

The same method was applied to obtain the pseudo-color $C_{\rm
F336W,F438W,F814W}$ width of the cluster AGB (RGB) stars, $W^{\rm
AGB\,(RGB),\,obs}_{C {\rm F336W,F438W,F814W}}$, as shown in panel (d), (e) and
(f) of Fig.~\ref{fig:wM15}. In this case we used the 20th percentile
distribution of the AGB (RGB) stars in the selected magnitude interval, as
shown in panel (d). Since the x-axis interval of the diagrams in panel (b),
(c), (e), and (f) spans 1\,mag, we can see that the observed width of the
cluster AGB (RGB) stars in the $C_{\rm F336W,F438W,F814W}$ is about a factor of
two smaller than that measured in the pseudo-color $C_{\rm F275W,F336W,F438W}$. 

The error in the determination of the width of the AGB and RGB stars, in both
the pseudo-colors, has been estimated by performing 10,000 bootstrapping tests
on random sampling with replacement. Each test was carried out by generating a
simulated sample containing sequential copies of the ${\rm
\Delta^{AGB\,(RGB),\,obs}}_{C\,{\rm F275W,F336W,F438W}}$ and ${\rm
\Delta^{AGB\,(RGB),\,obs}}_{C\,{\rm F336W,F438W,F814W}}$. Then a random
sub-sample composed of a number of stars equal to the observed one was
extracted and the resulting width computed.  The 68.27th percentile of the
distribution of simulated measurements was taken as the standard error of the
observed width.

The procedure described so far has been applied to measure the AGB and RGB
width of the other clusters. Since the number of detected AGB stars varies from
one cluster to another, as well as the covered F814W magnitude interval,
the number of adopted bins has been changed accordingly from a minimum
of 2 to a maximum of 3. The AGB members of \ngc6304 are oddly
distributed in magnitude, with the majority of them spanning a very narrow
interval. For this reason, it has not been possible to get a reliable estimate
of their pseudo-color extension, and the cluster has not been included in the
analysis described in the forthcoming section.

%%%%%%%%%%%%%%%%%%%%%%%%% FIGURE 12 %%%%%%%%%%%%%%%%%%%%%%%%%%%
%%%%%%%%%%%%%%%%%%%%%%%%%%%%%%%%%%%%%%%%%%%%%%%%%%%%%%%%%%%%%%%

Photometric errors also affect the observed pseudo-color spread by introducing
a spurious contribution to the intrinsic width of the cluster AGB and RGB
stars. To estimate this additional spread we took advantage of the photometric
errors previously determined, to simulate an artificial pseudo-color
distribution and measure the corresponding width, with the same procedure used
for the observations. The simulated width values have been subtracted in
quadrature from the corresponding observed width values to obtain the intrinsic
widths $W^{\rm AGB}_{C\,{\rm F275W,F336W,F438W}}$, $W^{\rm RGB}_{C\,{\rm
F275W,F336W,F438W}}$, $W^{\rm AGB}_{C\,{\rm F336W,F438W,F814W}}$ and $W^{\rm
RGB}_{C\,{\rm F336W,F438W,F814W}}$. We have reported the AGB and RGB intrinsic
width together with the corresponding error at columns 4 and 5, and 6 and 7,
respectively for the pseudo-color $C_{\rm F275W,F336W,F438W}$ and $C_{\rm
F275W,F336W,F438W}$. 

The listed $W^{\rm AGB}_{C\,{\rm F275W,F336W,F438W}}$ and $W^{\rm AGB}_{C\,{\rm
F336W,F438W,F814W}}$ values indicate that the intrinsic color spread of AGB
stars in all the clusters with more than 9 AGB members, is larger than that
expected from photometric errors alone. This finding, based on the largest
database of GCs analyzed so far, provides a clear indication that AGB sequences
of the 35 analyzed GCs host multiple stellar populations.

%%%%%%%%%%%%%%%%%%%%%%%%%%%%%%%%%%%%%%%%%%%%%%%%%%%%%%%%%%%%%%%%%%%%%%%%%%%%%%%%%%%%%%%%%%%%%%%%%%%%%%%%%%%%%%%%%%%%%
\begin{deluxetable*}{l*{2}{c}*{4}{C}}
\def\arraystretch{0.95}
\tabletypesize{\footnotesize}
%\tablewidth{10cm}
\tablecaption{Number of AGB members and bright RGB stars detected in each cluster, and corresponding intrinsic width and error, expressed in magnitudes, in the pseudo-color
$C_{\rm F275W,F336W,F438W}$ and $C_{\rm F275W,F336W,F438W}$.\label{tab1}}
\tablehead{
\colhead{ID} & \colhead{$\mathrm{N_{AGB}}$} & \colhead{$\mathrm{N_{RGB}}$} \ 
& \multicolumn1c{$W^{\rm AGB}_{C\,{\rm F275W,F336W,F438W}}$} & \multicolumn1c{$W^{\rm RGB}_{C\,{\rm F275W,F336W,F438W}}$} \
& \multicolumn1c{$W^{\rm AGB}_{C\,{\rm F336W,F438W,F814W}}$} & \multicolumn1c{$W^{\rm RGB}_{C\,{\rm F336W,F438W,F814W}}$}
}
\decimals
\startdata
\ngc104  &  29 &  71 & 0.400 \pm 0.023 & 0.439 \pm 0.074 & 0.136 \pm 0.023 & 0.156 \pm 0.036 \\
\ngc288  &   3 &   6 &   \textrm{-}    &   \textrm{-}    &   \textrm{-}    &   \textrm{-}    \\
\ngc362  &  29 &  87 & 0.235 \pm 0.051 & 0.299 \pm 0.028 & 0.120 \pm 0.031 & 0.116 \pm 0.007 \\
\ngc1261 &  31 &  69 & 0.288 \pm 0.039 & 0.305 \pm 0.014 & 0.099 \pm 0.016 & 0.118 \pm 0.010 \\
\ngc1851 &  38 &  75 & 0.375 \pm 0.031 & 0.359 \pm 0.044 & 0.140 \pm 0.018 & 0.129 \pm 0.024 \\
\ngc2298 &   7 &  15 &   \textrm{-}    &   \textrm{-}    &   \textrm{-}    &   \textrm{-}    \\
\ngc2419 &  83 & 373 & 0.333 \pm 0.025 & 0.596 \pm 0.076 & 0.109 \pm 0.020 & 0.142 \pm 0.007 \\
\ngc2808 &  83 & 196 & 0.411 \pm 0.034 & 0.464 \pm 0.020 & 0.183 \pm 0.015 & 0.136 \pm 0.011 \\
\ngc3201 &   3 &   4 &   \textrm{-}    &   \textrm{-}    &   \textrm{-}    &   \textrm{-}    \\
\ngc4590 &   7 &  15 &   \textrm{-}    &   \textrm{-}    &   \textrm{-}    &   \textrm{-}    \\
\ngc4833 &  16 &  50 & 0.188 \pm 0.025 & 0.304 \pm 0.027 & 0.087 \pm 0.032 & 0.099 \pm 0.012 \\
\ngc5024 &  32 & 112 & 0.197 \pm 0.022 & 0.227 \pm 0.012 & 0.066 \pm 0.016 & 0.074 \pm 0.010 \\
\ngc5053 &   0 &   - &   \textrm{-}    &   \textrm{-}    &   \textrm{-}    &   \textrm{-}    \\
\ngc5139 &  31 &  53 & 0.205 \pm 0.052 & 0.415 \pm 0.041 & 0.113 \pm 0.029 & 0.190 \pm 0.031 \\
\ngc5272 &  15 &  71 & 0.210 \pm 0.035 & 0.282 \pm 0.024 & 0.103 \pm 0.025 & 0.094 \pm 0.013 \\
\ngc5286 &  69 & 138 & 0.294 \pm 0.042 & 0.309 \pm 0.020 & 0.161 \pm 0.026 & 0.155 \pm 0.018 \\
\ngc5466 &   8 &  15 &   \textrm{-}    &   \textrm{-}    &   \textrm{-}    &   \textrm{-}    \\
\ngc5897 &   5 &  12 &   \textrm{-}    &   \textrm{-}    &   \textrm{-}    &   \textrm{-}    \\
\ngc5904 &  15 &  54 & 0.242 \pm 0.057 & 0.285 \pm 0.046 & 0.096 \pm 0.042 & 0.132 \pm 0.024 \\
\ngc5927 &  17 &  52 & 0.394 \pm 0.025 & 0.269 \pm 0.097 & 0.166 \pm 0.018 & 0.136 \pm 0.038 \\
\ngc5986 &  38 &  98 & 0.224 \pm 0.018 & 0.314 \pm 0.021 & 0.110 \pm 0.027 & 0.125 \pm 0.017 \\
\ngc6093 &  35 & 117 & 0.200 \pm 0.039 & 0.337 \pm 0.023 & 0.091 \pm 0.027 & 0.111 \pm 0.012 \\
\ngc6101 &  11 &  17 & 0.102 \pm 0.021 & 0.119 \pm 0.037 & 0.048 \pm 0.027 & 0.033 \pm 0.009 \\
\ngc6121 &   0 &   - &   \textrm{-}    &   \textrm{-}    &   \textrm{-}    &   \textrm{-}    \\
\ngc6144 &   7 &  12 &   \textrm{-}    &   \textrm{-}    &   \textrm{-}    &   \textrm{-}    \\
\ngc6171 &  13 &  16 & 0.345 \pm 0.081 & 0.277 \pm 0.060 & 0.134 \pm 0.047 & 0.118 \pm 0.012 \\
\ngc6205 &  14 &  66 & 0.256 \pm 0.030 & 0.336 \pm 0.036 & 0.078 \pm 0.020 & 0.117 \pm 0.014 \\
\ngc6218 &   4 &  21 &   \textrm{-}    &   \textrm{-}    &   \textrm{-}    &   \textrm{-}    \\
\ngc6254 &  10 &  25 & 0.355 \pm 0.056 & 0.327 \pm 0.049 & 0.155 \pm 0.034 & 0.114 \pm 0.023 \\
\ngc6304 &  11 &  28 &   \textrm{-}    &   \textrm{-}    &   \textrm{-}    &   \textrm{-}    \\
\ngc6341 &  15 &  64 & 0.288 \pm 0.069 & 0.173 \pm 0.026 & 0.218 \pm 0.063 & 0.066 \pm 0.011 \\
\ngc6352 &   4 &  11 &   \textrm{-}    &   \textrm{-}    &   \textrm{-}    &   \textrm{-}    \\
\ngc6362 &   5 &  14 &   \textrm{-}    &   \textrm{-}    &   \textrm{-}    &   \textrm{-}    \\
\ngc6366 &   2 &   3 &   \textrm{-}    &   \textrm{-}    &   \textrm{-}    &   \textrm{-}    \\
\ngc6388 & 158 & 268 & 0.522 \pm 0.027 & 0.443 \pm 0.021 & 0.291 \pm 0.042 & 0.211 \pm 0.025 \\
\ngc6397 &   1 &   2 &   \textrm{-}    &   \textrm{-}    &   \textrm{-}    &   \textrm{-}    \\
\ngc6441 & 161 & 246 & 0.540 \pm 0.026 & 0.608 \pm 0.047 & 0.199 \pm 0.026 & 0.216 \pm 0.040 \\
\ngc6496 &   2 &  11 &   \textrm{-}    &   \textrm{-}    &   \textrm{-}    &   \textrm{-}    \\
\ngc6535 &   3 &   9 &   \textrm{-}    &   \textrm{-}    &   \textrm{-}    &   \textrm{-}    \\
\ngc6541 &   8 &  27 &   \textrm{-}    &   \textrm{-}    &   \textrm{-}    &   \textrm{-}    \\
\ngc6584 &  11 &  27 & 0.198 \pm 0.044 & 0.212 \pm 0.021 & 0.135 \pm 0.053 & 0.107 \pm 0.021 \\
\ngc6624 &  15 &  33 & 0.415 \pm 0.069 & 0.444 \pm 0.044 & 0.138 \pm 0.025 & 0.164 \pm 0.020 \\
\ngc6637 &  29 &  54 & 0.367 \pm 0.031 & 0.311 \pm 0.026 & 0.133 \pm 0.015 & 0.129 \pm 0.009 \\
\ngc6652 &   8 &  15 &   \textrm{-}    &   \textrm{-}    &   \textrm{-}    &   \textrm{-}    \\
\ngc6656 &  10 &  70 & 0.264 \pm 0.036 & 0.299 \pm 0.033 & 0.161 \pm 0.038 & 0.178 \pm 0.021 \\
\ngc6681 &  12 &  23 & 0.293 \pm 0.063 & 0.381 \pm 0.056 & 0.140 \pm 0.035 & 0.154 \pm 0.026 \\
\ngc6715 &  84 & 248 & 0.327 \pm 0.037 & 0.431 \pm 0.022 & 0.160 \pm 0.038 & 0.146 \pm 0.007 \\
\ngc6717 &   7 &  14 &   \textrm{-}    &   \textrm{-}    &   \textrm{-}    &   \textrm{-}    \\
\ngc6723 &  23 &  39 & 0.391 \pm 0.047 & 0.457 \pm 0.028 & 0.149 \pm 0.030 & 0.128 \pm 0.015 \\
\ngc6752 &  11 &  29 & 0.231 \pm 0.051 & 0.435 \pm 0.050 & 0.364 \pm 0.072 & 0.142 \pm 0.026 \\
\ngc6779 &  14 &  27 & 0.155 \pm 0.022 & 0.274 \pm 0.052 & 0.082 \pm 0.024 & 0.078 \pm 0.023 \\
\ngc6809 &   3 &  10 &   \textrm{-}    &   \textrm{-}    &   \textrm{-}    &   \textrm{-}    \\
\ngc6838 &   6 &   6 &   \textrm{-}    &   \textrm{-}    &   \textrm{-}    &   \textrm{-}    \\
\ngc6934 &  23 &  56 & 0.301 \pm 0.057 & 0.329 \pm 0.034 & 0.128 \pm 0.029 & 0.117 \pm 0.014 \\
\ngc6981 &  14 &  22 & 0.260 \pm 0.074 & 0.313 \pm 0.082 & 0.085 \pm 0.045 & 0.109 \pm 0.014 \\
\ngc7078 &  34 & 192 & 0.224 \pm 0.037 & 0.257 \pm 0.009 & 0.149 \pm 0.039 & 0.104 \pm 0.005 \\
\ngc7089 &  68 & 216 & 0.267 \pm 0.021 & 0.303 \pm 0.017 & 0.114 \pm 0.011 & 0.124 \pm 0.009 \\
\ngc7099 &   6 &  14 &   \textrm{-}    &   \textrm{-}    &   \textrm{-}    &   \textrm{-}    \\
\enddata
\end{deluxetable*}
%%%%%%%%%%%%%%%%%%%%%%%%%%%%%%%%%%%%%%%%%%%%%%%%%%%%%%%%%%%%%%%%%%%%%%%%%%%%%%%%%%%%%%%%%%%%%%%%%%%%%%%%%%%%%%%%%%

\section{Impact of light elements on the colors of AGB stars}\label{sec:teo}

To qualitatively investigate the typical effects of changing the abundances of
He, C, N, and O on the colors and magnitudes of AGB stars, we combined
isochrones and synthetic spectra of AGB stars with appropriate chemical
compositions.  We used the evolutionary code ATON 2.0 \citep[][]{ventura98a,
mazzitelli99,ventura09a} to calculate 12\,Gyr-old isochrones with
\feh$=-1.5$, \afe$=0.4$ but different RGB mass loss, helium content and C,
N, and O abundance \citep[see][for details]{dantona02a,tailo19,tailo20}.  

We first investigated the effect of helium and mass-loss variations on the
AGBs.  We calculated an isochrone with pristine helium content (Y$=0.25$), 
[C/Fe]=0.0, [N/Fe]=0.0 and [O/Fe]=0.4 and RGB mass loss $\mu=0.30 M_{\odot}$.
Moreover, we derived two helium-enhanced isochrones with Y$=0.28$ with the same
C, N and O content as above and different mass losses of $\mu=0.30 M_{\odot}$
and $\mu=0.45 M_{\odot}$.  Finally, we computed three isochrones with Y$=0.28$,
[C/Fe]=$-$0.5, [N/Fe]=1.21 and [O/Fe]=$-$0.1 and $\mu = 0.30, 0.45, 0.60
M_{\odot}$.

To account for the effect of C, N and O variations on the studied isochrones,
we extended the procedure by \citet{milone12a,milone18b} to the AGB. We
identified five points along each isochrone with [C/Fe]=0, [N/Fe]=0 and
[O/Fe]=0.4 and extracted their \teff\ and gravity ($g$).  We used ATLAS12 and
SYNTHE computer programs
\citep[e.g.][]{castelli05,kurucz05,sbordone07,sbordone11} to calculate a
spectrum with [C/Fe]=0.0, [N/Fe]=0.0, and [O/Fe]=0.4 (reference spectrum) and a
spectrum with enhanced nitrogen [N/Fe]=1.21 and depleted carbon and oxygen
([C/Fe]$=-0.5$ and [O/Fe]$=-0.1$). All spectra are computed over the wavelength
interval between 2000\,\AA\ and 10,000\,\AA\ and are convoluted with the
throughput of the F275W, F336W, F438W WFC3/UVIS filters and the F606W and F814W
ACS/WFC filters used in this paper.

\begin{figure}
\centering
\includegraphics[height=8.0cm,clip,trim={0.8cm 7.7cm 0.3cm 4cm}]{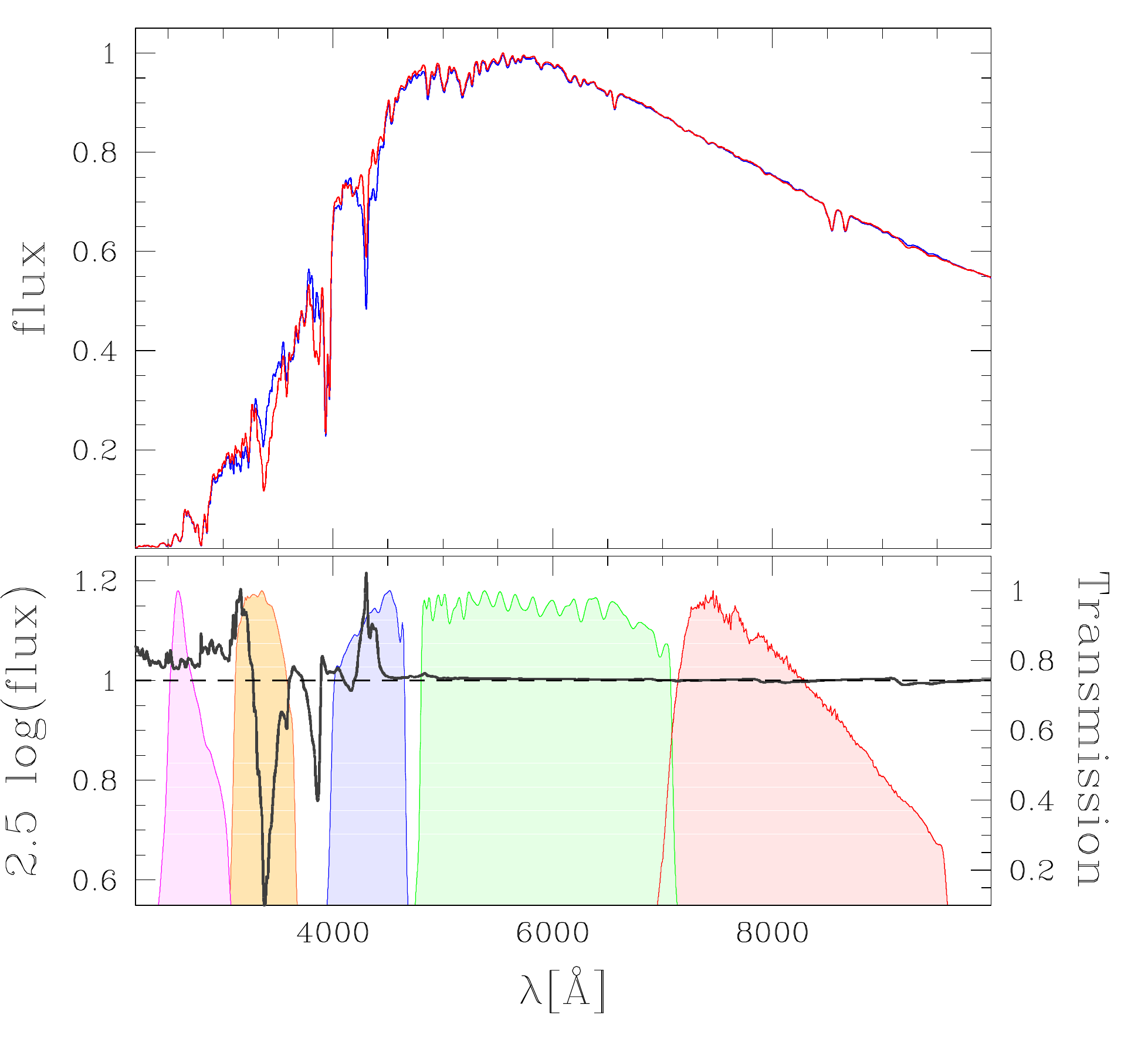} 
\includegraphics[height=5.6cm,clip,trim={0.8cm 11.5cm 0.3cm 4.5cm}]{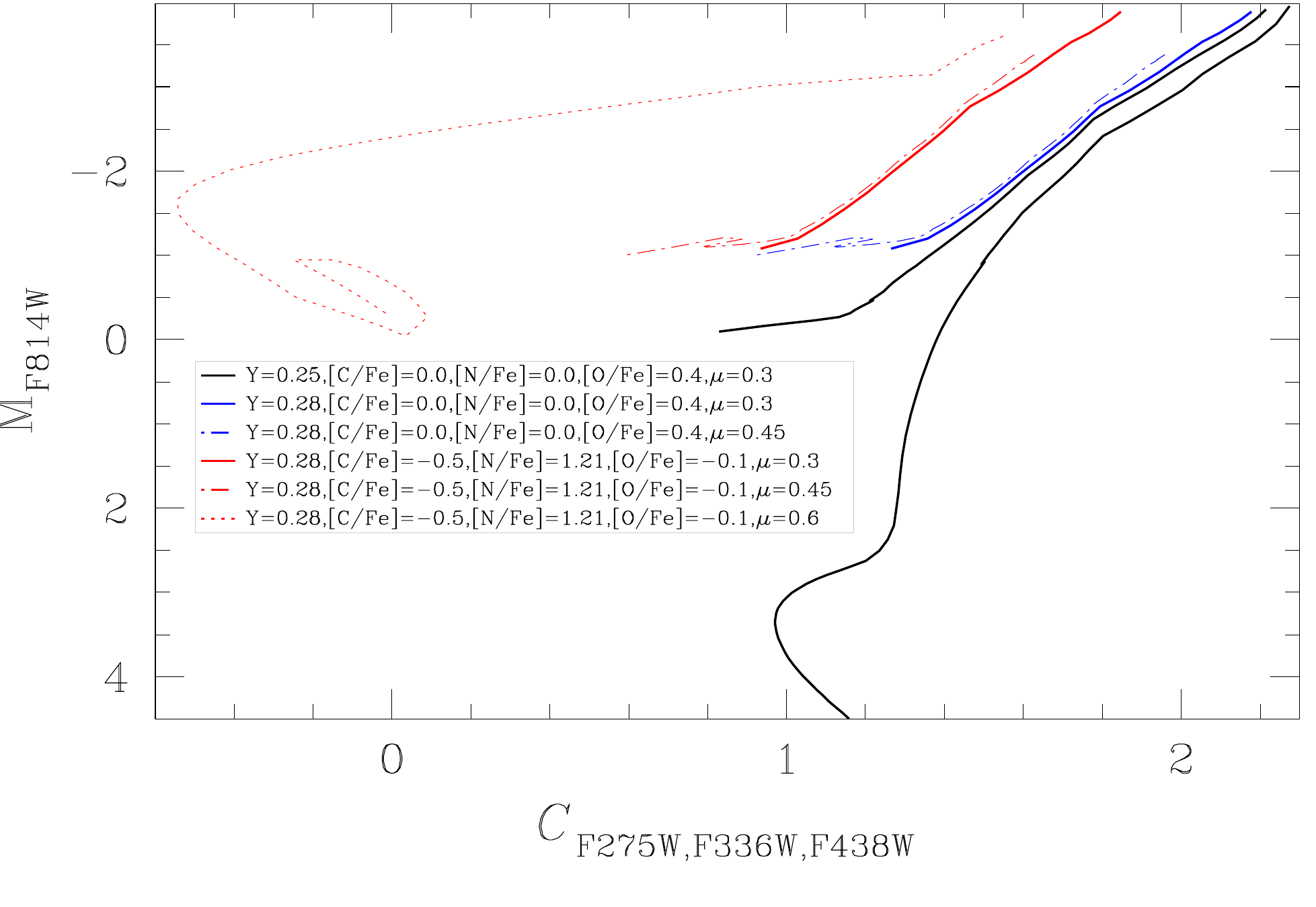} 
\includegraphics[height=5.6cm,clip,trim={0.8cm 11.5cm 0.3cm 4.5cm}]{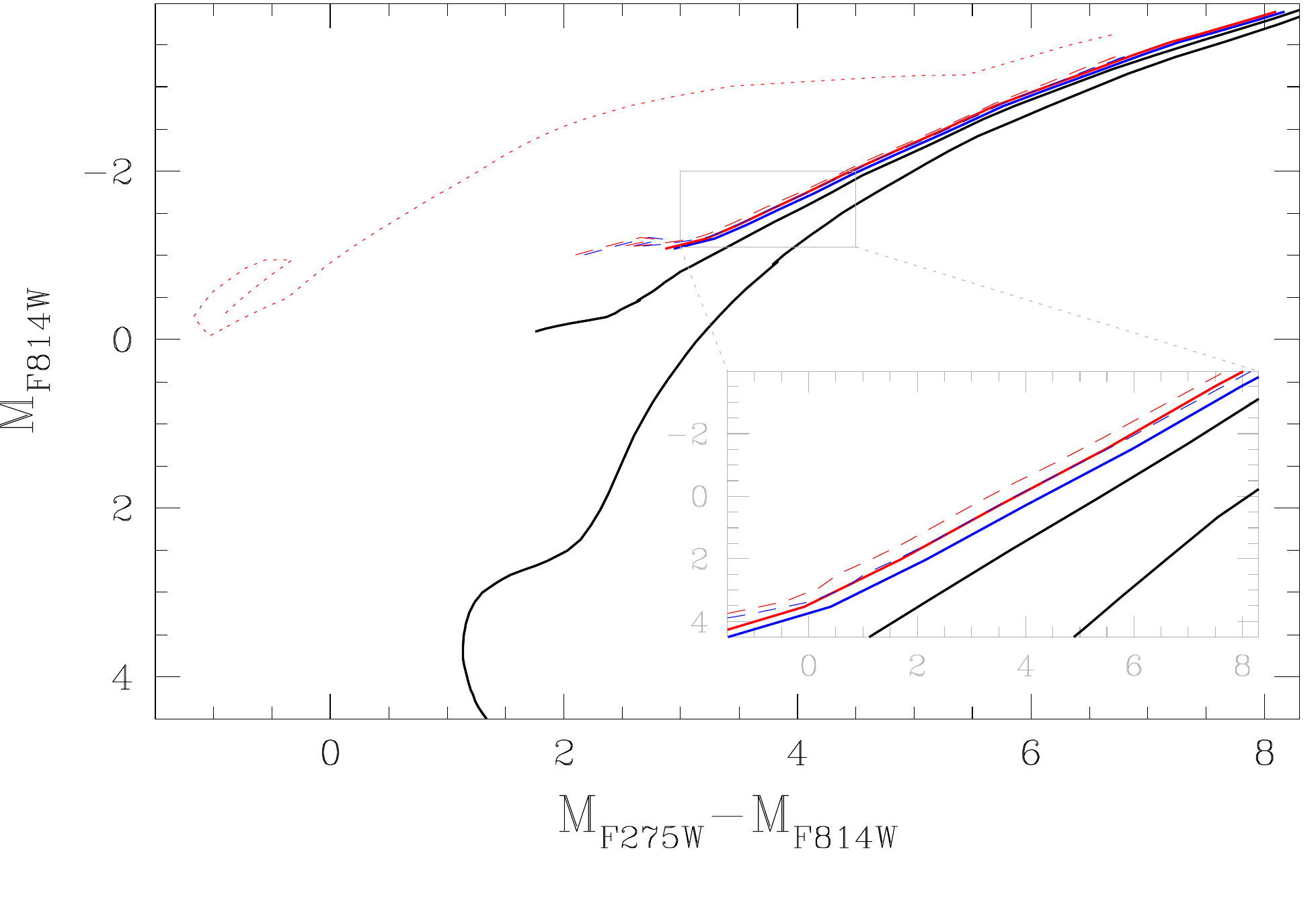} 
\caption{\textit{Upper panels}: Comparison between the spectra of AGB stars
	with the same atmospheric parameters but chemical composition typical
	of 1G (red spectrum) and 2G stars (blue spectrum). The magnitude
	difference between the two spectra is plotted as a function of the
	wavelength in the bottom, where we also show the normalized
	transmission curves of the F275W (magenta), F336W (orange), F438W
	(blue), F606W (green) and F814W (red) filters used in this paper.
	\textit{Middle and lower panel}: isochrones of a 13\,Gyr-old stellar
	population with \feh=$-1.5$ and \afe=0.4 (black isochrone) in the
	$M_{\rm F814W}$ vs. ${C_{\rm F275W,F336W,F438W}}$ and
	$M_{\rm F814W}$ vs. ${M_{\rm F275W}-M_{\rm F814W}}$ CMDs.  The
	colored isochrone correspond to AGB stars with different chemical
	composition and mass loss, as reported in the middle panel.\label{fig:teo1}}
\end{figure}

As an example, the uppermost panel of Figure~\ref{fig:teo1} compares the spectra of
two AGB stars with Y=0.25, $\mu=0.30 M_{\odot}$, $T_{\rm eff}=4875$K and
$\log{g}=1.76$. The red spectrum is representative of 1G star and has
[C/Fe]=0.0, [N/Fe]=0.0 and [O/Fe]=0.4, whereas the C, N, O content of the blue
spectrum resembles 2G stars ([C/Fe]=$-$0.5, [N/Fe]=1.21 and [O/Fe]=$-$0.1). We
calculate the magnitude difference between each point of the two spectra as
-2.5 times the logarithm of their flux ratio and plot this quantity as a
function of the wavelength. For comparison, we over-plotted the transmission
curves of the five filters used in this paper. Clearly, the 2G spectrum
provides F336W magnitude fainter than the 1G ones, mostly due to the NH band
around 3,600\,\AA. 2G stars also exhibit brighter F275W and F438W magnitudes,
as a consequence of the strength of the OH and CH molecular bands,
respectively. Hence, C, N, and O variations in the spectra of AGB stars show a
qualitative behavior similar to that observed in RGB stars \citep[e.g.][]{milone12a}. 

Synthetic spectra are convoluted with the throughput of the five filters used
in this paper to derive the corresponding magnitudes. The magnitude differences
between the comparison and the reference spectrum are added to the
corresponding magnitudes of the isochrones with 1G-like chemical composition to
derive the corresponding isochrones enhanced in N and depleted in C and O. 

%%%%%%%%%%%%%%%%%%%%%%%%% FIGURE 13 %%%%%%%%%%%%%%%%%%%%%%%%%%%
%%%%%%%%%%%%%%%%%%%%%%%%%%%%%%%%%%%%%%%%%%%%%%%%%%%%%%%%%%%%%%%

\section{Comparison between AGB and RGB width\label{sec:AGBvs}}
The photometric width as standard measurement of the amount of
internal chemical variations gives us the ability to compare the properties of
MPs across the entire parameter-space spanned by GCs. In this regard, recent
works have demonstrated that a significant correlation exists between the spread of
RGB in the indices $C_{\rm F275W,F336W,F438W}$ and $C_{\rm F336W,F438W,F814W}$,
and cluster's metallicity and mass \citep{milone17,lagioia19a}.  
For this reason we decided to study the relation between the AGB and RGB width.

\begin{figure*}
\centering
\includegraphics[width=\textwidth]{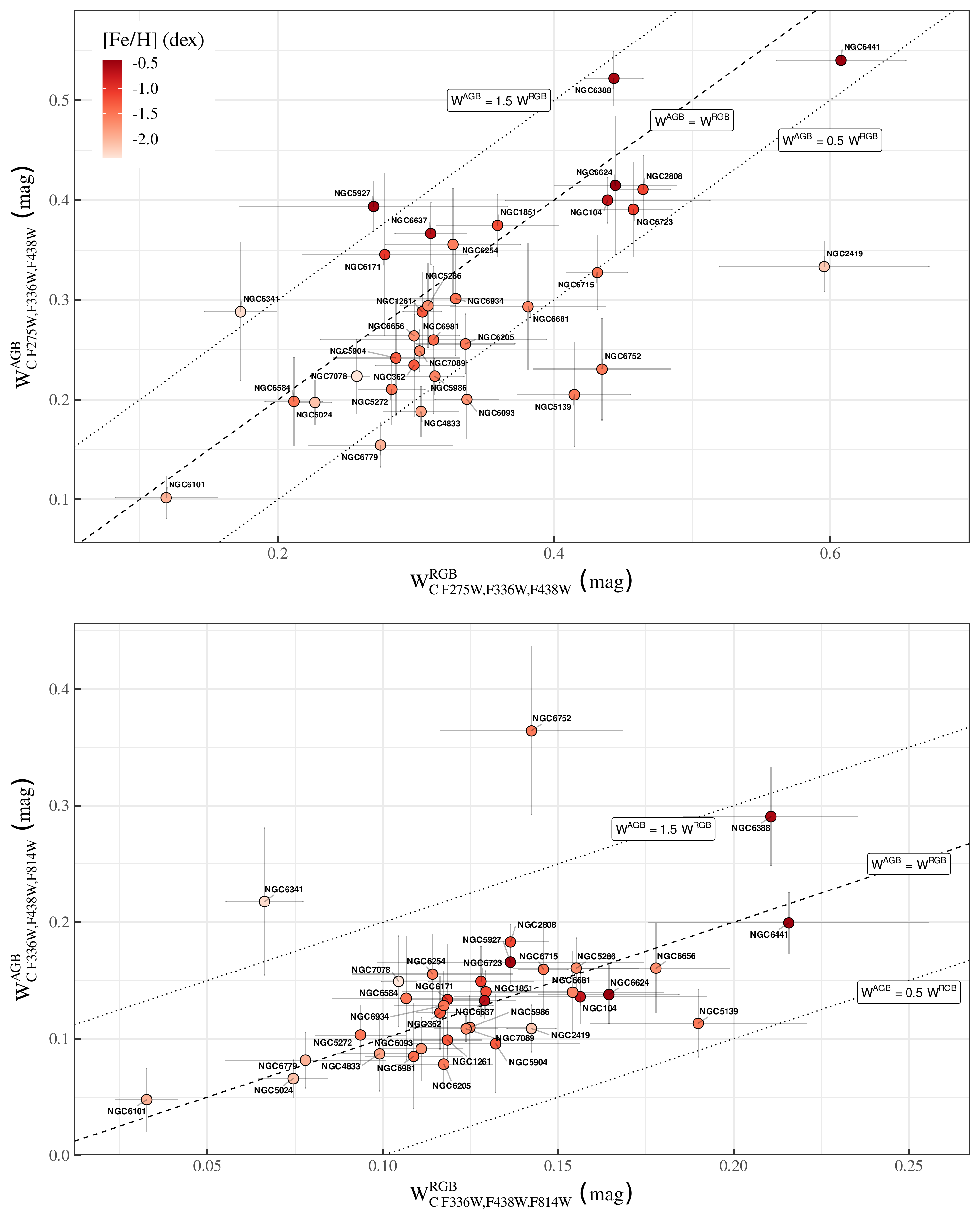} 
\caption{\textit{Top}: $W^{\rm AGB}_{C\,{\rm F275W,F336W,F438W}}$ vs.
	$W^{\rm RGB}_{C\,{\rm F275W,F336W,F438W}}$ of the 35 GCs in
	Table~\ref{tab1}. The color of each point maps the cluster's
	metallicity, according to the 
	scale reported in the legend. \textit{Bottom}: $W^\mathrm{AGB}_{C\,{\rm
	F336W,F438W,F814W}}$ vs. $W^{\rm RGB}_{C\,{\rm F336W,F438W,F814W}}$
	of the same clusters. The reference dashed and dotted lines overplotted
	in each panel represent the relations indicated by the corresponding labels.\label{fig:w35GCs}}
\end{figure*}

In the top panel of Figure~\ref{fig:w35GCs}, we plot $W^{\rm AGB}_{C\,{\rm
F275W,F336W,F438W}}$ vs. $W^{\rm RGB}_{C\,{\rm F275W,F336W,F438W}}$ of the 35
clusters, listed in Table~\ref{tab1}, for which the measurement of the relative
quantities is available. In the diagram, each point has been color-coded
according to the corresponding cluster's metallicity \feh\ \citep[][2010
update]{harris96a}, mapped to the color scale reported in the legend. We see
that the two represented quantities follow, on average, a positive correlation.
In order to provide a reference for the comparison, we overplotted three lines
representing the identity relation $W^{\rm AGB} = W^{\rm RGB}$ (dashed
line), and the relations $W^{\rm AGB} = 1.5\,W^{\rm RGB}$ and
$W^{\rm AGB} = 0.5\,W^{\rm RGB}$ (top and bottom dotted lines). We
observe that the majority of GCs (27 out of 35 GCs) have AGB width smaller than
the RGB width, and that among them, 7 GCs have $W^{\rm AGB} < 0.5
W^{\rm RGB}$, with the remarkable cases of \ngc2419, \ngc5139 (\ome), and \ngc6752, 
the three GCs with the smallest AGB/RGB width ratio. On the other
hand, \ngc6341 and \ngc5927 are the only two clusters lying above the
$W^{\rm AGB} = 0.5\,W^{\rm RGB}$ line. Their relatively large error bars,
however, do not allow to draw strong conclusions.  The color distribution seems
to suggest the presence of a mild monotonic trend between the AGB intrinsic
width and metallicity, with the most metal-rich clusters having on average
larger width values.  

The bottom panel displays the scatter plot of $W^{\rm AGB}_{C\,{\rm
F336W,F438W,F814W}}$ vs. $W^{\rm RGB}_{C\,{\rm F336W,F438W,F814W}}$. As before,
we also overplotted the same three reference lines. We observe that all the
clusters lie within the region defined by the $W^{\rm AGB} = 1.5\,W^{\rm RGB}$
and $W^{\rm AGB} = 0.5\,W^{\rm RGB}$ lines, with the clusters quite uniformly
distributed above and below the identity relation line.  Only two clusters,
namely \ngc6341 and \ngc6752, have AGB/RGB width ratios larger than 1.5. While
the first cluster shows a behavior similar to that observed in the previous
case, that of \ngc6752 is reversed. However, we highlight the fact that the AGB
width estimate of this cluster has been derived from a relatively low number of
stars (11) and, as a consequence, it is poorly constrained, as also indicated
by its large error bars. Therefore, again, no firm conclusion can be achieved
about these two GCs.  We finally notice the absence of any trend with
metallicity in this color combination.

The relation between AGB and RGB width visible in both the diagrams clearly
indicates that none of the observed clusters have a mono-populated AGB
sequence. The trend visible in the top panel shows that, on average, the AGB width is
smaller than the RGB width. In the hypothesis that in each cluster all the
stellar populations in the RGB evolve to the AGB, we would expect to observe AGB
widths comparable with or larger than those of the corresponding RGB stars. The
displayed trend instead, indicates that the AGB width is on average smaller
than the RGB width. This finding clearly suggests that in the majority of
clusters a significant fraction of 2G stars does not evolve to the AGB, as
predicted by the AGB-\textit{manqu\'e} scenario. Our finding, based on a large
sample of GCs, provides a robust evidence of the existence of this phenomenon.

We emphasize that the displayed trend may include some spurious scatter due to
the specific evolutionary conditions of each different cluster: mass loss has,
indeed, an important effect on the color spread of AGB stars, as seen in the
previous section. This implies that a precise evaluation of the chemical
variations corresponding to every AGB width measurement can be obtained through
the comparison between observations and theoretical models, suitably tailored
to the specific properties of each analyzed cluster. However, a one-to-one
comparison between the AGB and RGB width of each cluster is beyond the purpose
of the present analysis.

%%%%%%%%%%%%%%%%%%%%%%%%% FIGURE 14 %%%%%%%%%%%%%%%%%%%%%%%%%%%
%%%%%%%%%%%%%%%%%%%%%%%%%%%%%%%%%%%%%%%%%%%%%%%%%%%%%%%%%%%%%%%

\section{The chromosome maps of AGB stars}\label{sec:AGBChM}
\begin{figure*}
\centering
\includegraphics[angle=270,width=\textwidth,trim=0cm 0cm 0cm 2cm,clip]{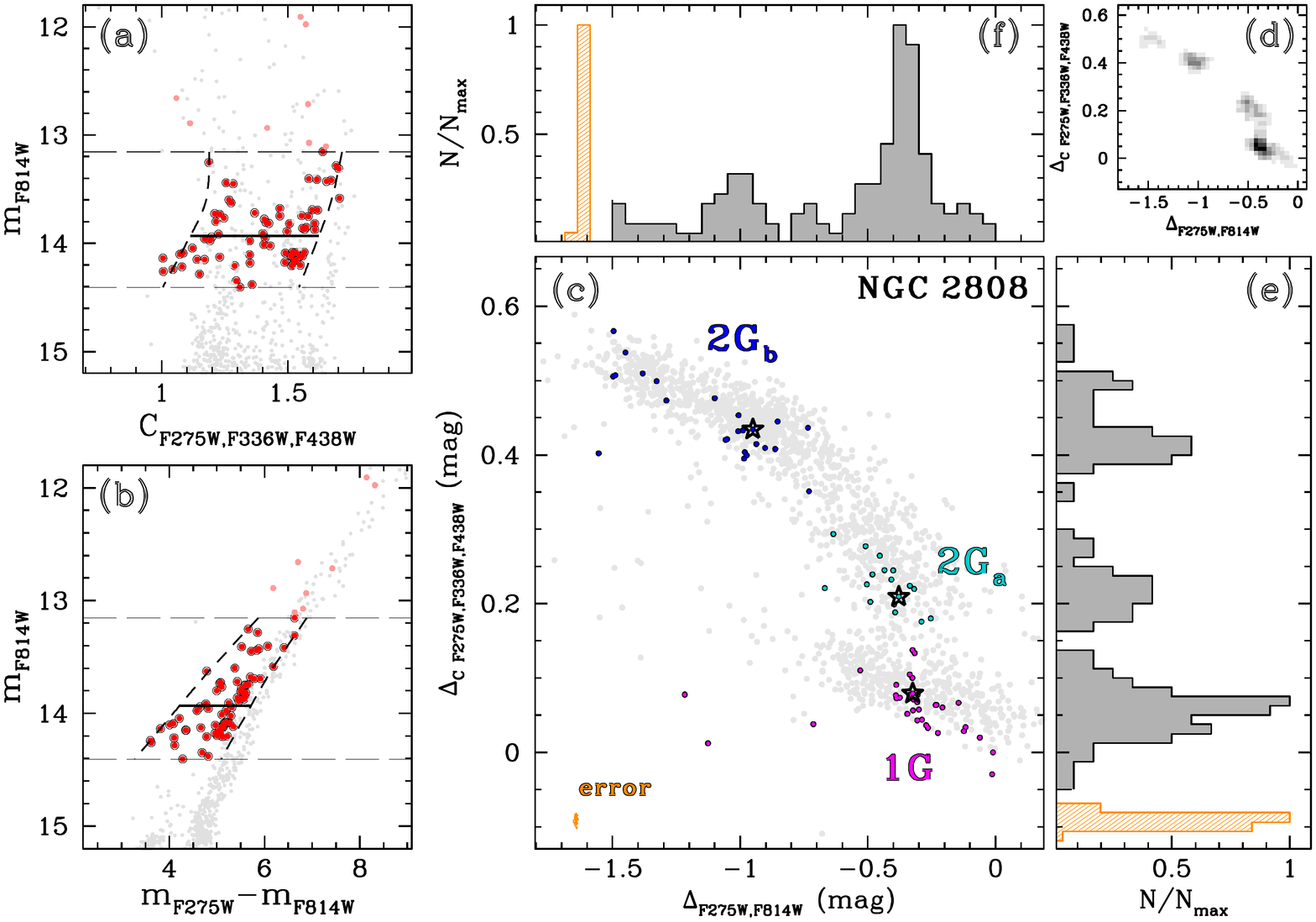} 
\caption{Chromosome map of \ngc2808 AGB stars. \textit{Panel (a)}: $m_{\rm
	F814W}$ vs. $C_{\rm F275W,F336W,F438W}$ CMD of the brightest cluster
	stars. AGB stars are represented as red points. AGB members excluded
	from the analysis are color-coded in light red.  Left and right dashed
	black curves represent, respectively, the 4th and 96th percentile of
	the pseudo-color distribution of the AGB stars in the selected
	magnitude interval, delimited by the two long-dashed horizontal lines.
	The horizontal solid black line outlines the 4th-96th pseudo-color
	width of the AGB stars. \textit{Panel (b)}: same as in panel (a) but in
	the $m_{\rm F814W}$ vs. $m_{\rm F275W}-m_{\rm F814W}$ CMD.
	\textit{Panel (c)}: ChM of the cluster AGB stars. The labels mark the
	different populations of AGB stars with the corresponding color. Black
	starred symbols identify the spectroscopic targets by \citet{marino17}.
	Background gray points mark the scaled-up ChM of the cluster RGB stars
	\citep{milone17} (see text for details). Orange points mark the ChM of
	the photometric errors.  \textit{Panel (d)}: density map diagram of the
	cluster AGB ChM.  \textit{Panel (e), (f)}: histogram distribution of
	the ChM along the vertical and horizontal axis, respectively. The
	orange-shaded histogram corresponds to the histogram distribution of
	the ChM of the photometric error.\label{fig:chm2808}} 
\end{figure*}

The analysis of the AGB width presented in the previous section provides a
strong indication about the presence of MPs along the AGB of the analyzed GCs,
as well as of the partial depletion of MPs due to the evolution of the most
chemically-enriched stars as AGB-\textit{manqu\'{e}}. To further investigate
this phenomenon, we exploited the ChM of AGB stars. The construction of the
ChM of the GCs analyzed in the present work, shown in Figure~\ref{fig:chm2808} for
the template cluster \ngc2808, is based on the method introduced by
\citet{milone15b} and extended to the AGB by \citet{marino17}. In the figure,
panel (a) displays the portion of the $m_{\rm F814W}$ vs. $C_{\rm F275W,F336W,F438W}$
pseudo-CMD of the cluster, zoomed on the AGB stars. AGB stars are represented as red
points, while all the other visible stars, mainly composed by bright RGBs, as
light-gray points.  As described in Section~\ref{sec:width}, we excluded the
brightest AGB cluster stars, which are shown as light-red points. Long-dashed
horizontal lines delimit the selected luminosity interval.  The left and right
dashed black curves represent, respectively, the 4th and 96th percentile of the
pseudo color-distribution of the AGB stars. They have been obtained by dividing
the entire luminosity interval in 4 bins and then computing, for each bin, the
color corresponding to the 4th and 96th percentile of the color distribution.
We then applied a boxcar-average smoothing to the observed distribution points
and finally fit the resulting pseudo-colors by a cubic spline along the
central magnitude of the bins. We then performed a double-normalization by
subtracting to the pseudo-color of each star, the corresponding value along the
96th percentile fit curve, and dividing the result by the distance between the
point of the 4th and 96th fit curve at the same magnitude \citep[see equation 1
and 2 in][]{milone17}. The ratio has been finally multiplied by the factor
$W^\mathrm{96th-4th, AGB}_{C {\rm F275W,F336W,F814W}}$, corresponding to the
intrinsic width of the AGB stars. This quantity is defined as the observed distance between
the fit curves at a reference luminosity equal to 5\,mag brighter than the
main-sequence turn-off in F814W band, minus the contribution due to the
photometric error, as described in the previous section. 

The luminosity of the main sequence turn-off has been determined by applying
the naive estimator method \citep{silverman86}. This procedure consists in
subdividing a pre-selected F814W magnitude range centered on the turn-off
region of the CMD, in a number of magnitude bins. For each bin the median color
and magnitude of stars is computed. The same algorithm is applied in sequence
by shifting the first bin magnitude by an fraction of the bin width. The
magnitude and color of all median points are then boxcar-average smoothed.
Finally the bluest color of the interpolating function is taken as the
$m^{MSTO}_{F814W}$ value. We found $W^{\rm 4th-96th, AGB}_{C\,{\rm
F275W,F336W,F438W}} = 0.508$\,mag. The result of the double normalization is a
new quantity called ${\rm \Delta}_{C\,{\rm F275W,F336W,F438W}}$. The same
procedure
has been applied to the $m_{\rm F814W}$ vs. $m_{\rm F275W}-m_{\rm F814W}$ CMD displayed in
panel (b). We found in this case $W^{\rm 4th-96th, AGB}_{\rm F275W,F814W} =
1.487$\,mag and obtained the new quantity ${\rm \Delta}_{\rm F275W,F814W}$.

Panel (c) displays the ChM, namely the ${\rm \Delta}_{C\,{\rm
F275W,F336W,F438W}}$ vs. ${\rm \Delta}_{\rm F275W,F814W}$ diagram of the AGB
stars of \ngc2808. A glance at the distribution of the stars immediately shows
that in this cluster, the AGB stars are not evenly distributed but rather
clustered in at least four distinct groups \citep[see][for
comparison]{marino17}. In order to understand if the observed scatter can be
accounted to the photometric error, we derived the ChM expected for a
mono-populated cluster. In this case the scatter is mostly due to the
observational errors. The center of the resulting distribution has been
arbitrarily shifted at coordinates $({\rm \Delta}_{\rm F275W,F814W}, {\rm
\Delta}_{C\,{\rm F275W,F336W,F438W}})$ = $\sim (-1.65,-0.85)$. It appears as an
ellipse flattened along the horizontal direction, with a major semi-axis of
$\approx 0.025$\,mag. We see that the vertical extension of the AGB ChM is
$\sim 0.1$\,mag, therefore about four times larger than that of the error
distribution. 

The density diagram plotted in panel (d) further confirms the presence of
discrete stellar populations.  It shows indeed two main groups of stars at
coordinates $\approx (-0.30, 0.15)$ and $\approx (-1.00, 0.15)$, and two other
less populated groups at $\approx (-0.40, 0.25)$  and $\approx (-1.40, 0.30)$.
With the aim of properly assigning a membership to each star in the various
groups, we built the histogram distribution of AGB stars along the y-axis and
x-axis. The results are shown in panel (e) and (f), respectively.  Each
histogram has been obtained by splitting the horizontal and vertical intervals
in bins of $0.10$\,mag, and computing the number of clusters falling in each
bin at steps equal to half bin width. This choice mitigates the arbitrary
decision of a given bin width.  In the histogram of panel (e), we can recognize
three distinct sections, centered at ${\rm \Delta}_{C\,{\rm
F275W,F336W,F438W}}$ = $\approx 0.06, 0.21, 0.40$\,mag. The histogram of the
last section shows a secondary split at $\approx 0.38$\,mag, in agreement with
the density map in panel (d).  The histogram in panel (f) shows two main peaks
at ${\rm \Delta}_{\rm F275W,F814W}$ = $\approx -0.30$\,mag and $\approx
-0.95$\,mag. Therefore, we used the vertical histogram distribution to
subdivide the ChM into three groups, that we named 1G (magenta points), ${\rm
2G_a}$ (azure points) and ${\rm 2G_b}$ (blue points). The nomenclature follows
the subdivision adopted by \citet{marino17}. For the sake of clarity, we used a
different color code for the three groups of stars. We finally cross-matched
the three AGB spectroscopic targets by \citet{marino17}, and marked them with a
black starred symbol in the ChM of panel (c). We see that these stars, which
have different O and Na abundance, belong to the different groups identified in
the AGB ChM, thus providing the definitive confirmation that 1G, ${\rm 2G_a}$,
and ${\rm 2G_b}$, correspond to distint stellar populations with different
light-element abundance

A detailed interpretation of the MP composition of the cluster AGB star can be
obtained by comparing the present ChM with that obtained from the analysis of
RGB stars \citep{milone17}. Since ChM are built from the measurement of the
spread of stars in specific color combinations, such a comparison is ultimately
connected to that of the widths. However, as already mentioned before, this
implies  a precise evaluation of the factors affecting the evolutionary path of
post-HB stars, which is not the purpose of the present analysis.
Notwithstanding, we decided to perform a crude comparison by assuming that the
effect of the presence of 1G and 2G stars in the extension of the ChM of AGB
and RGB stars is comparable. Therefore, we decided to scale-up the width of the
RGB stars in $m_{\rm F275W}-m_{\rm F814W}$ and $C_{\rm F275W,F336W,F438W}$
\citep[see][Table~2]{milone17} to that of the AGB stars derived in this work.
The scaled RGB ChM is represented by the gray points plotted on background in
panel (c).

Interestingly, we observe that the AGB ChM broadly reproduces the overdensities
corresponding to the sub-populations observed in the RGB ChM.  In particular we
observe a similar elongation between the AGB and RGB 1G populations, with three
AGB stars (${\rm \Delta}_{\rm F275W,F814W} \lesssim -0.5$\,mag) attaining ${\rm
\Delta}_{C\,{\rm F275W,F336W,F438W}}$ values bluer than the blue ${\rm
\Delta}_{\rm F275W,F814W}$ tail of the RGB stars. On the other side ${\rm
2G_a}$ and ${\rm 2G_b}$ AGB stars appear to follow the same trend of the
sub-groups C and D of the 2G RGB population \citep[see][]{marino17}.  We also
notice the presence of a blue ${\rm \Delta}_{\rm F275W,F814W}$ tail in the
${\rm 2G_b}$ group, that overlaps the location of the extreme helium-enhanced
cluster (E) population \citep[$\mathrm{\delta\,Y} \geq 0.1$][]{milone15b}.
This appears to be in contrast with the predictions of evolutionary models of
very helium-rich stars, which must avoid the AGB phase
\citep[e.g.][]{chantereau16}. Of course, spectroscopic investigation is
mandatory to understand whether these AGB stars are the counterparts of RGB and
MS stars with extreme helium abundances, or if they have intermediate helium
content but have experienced severe RGB mass loss. 

For reference we also computed the histogram distribution
of the ChM of the photometric errors. The corresponding orange-shaded
histograms appears to be spread over an interval visibly narrower than the
sub-structures visible in both the ChM histogram distribution in panel (e) and
(f).

%%%%%%%%%%%%%%%%%%%%%%%%% FIGURE 15 %%%%%%%%%%%%%%%%%%%%%%%%%%%
%%%%%%%%%%%%%%%%%%%%%%%%%%%%%%%%%%%%%%%%%%%%%%%%%%%%%%%%%%%%%%%

We used the same procedure to build the ChM of all the other GCs with
$\mathrm{N_{AGB}} > 25$ in our database (see column 2 of Table~\ref{tab1}).
The heterogeneous F814W magnitude distribution of the AGB stars of \ngc104 and
\ngc6637, in the CMDs used to build their ChM, did not allow to obtain reliable
estimate of the AGB widths. For this reason, these two GCs have been excluded
from the following analysis.

\begin{figure*}
\centering
\includegraphics[width=0.7\textwidth,trim=.7cm 6cm  .2cm 4.5cm,clip]{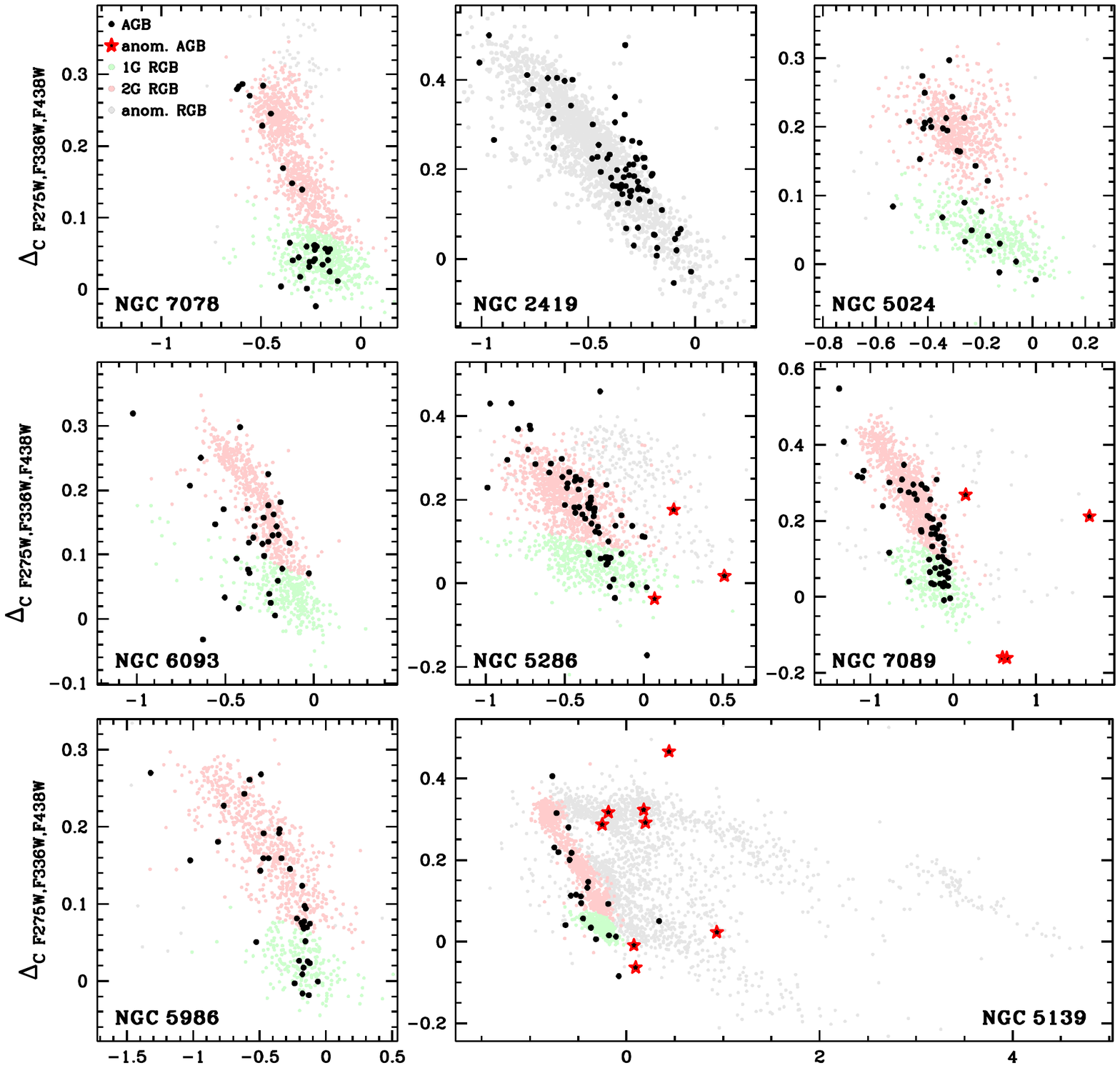} 
\includegraphics[width=0.7\textwidth,trim=.7cm 11cm .2cm 4.5cm,clip]{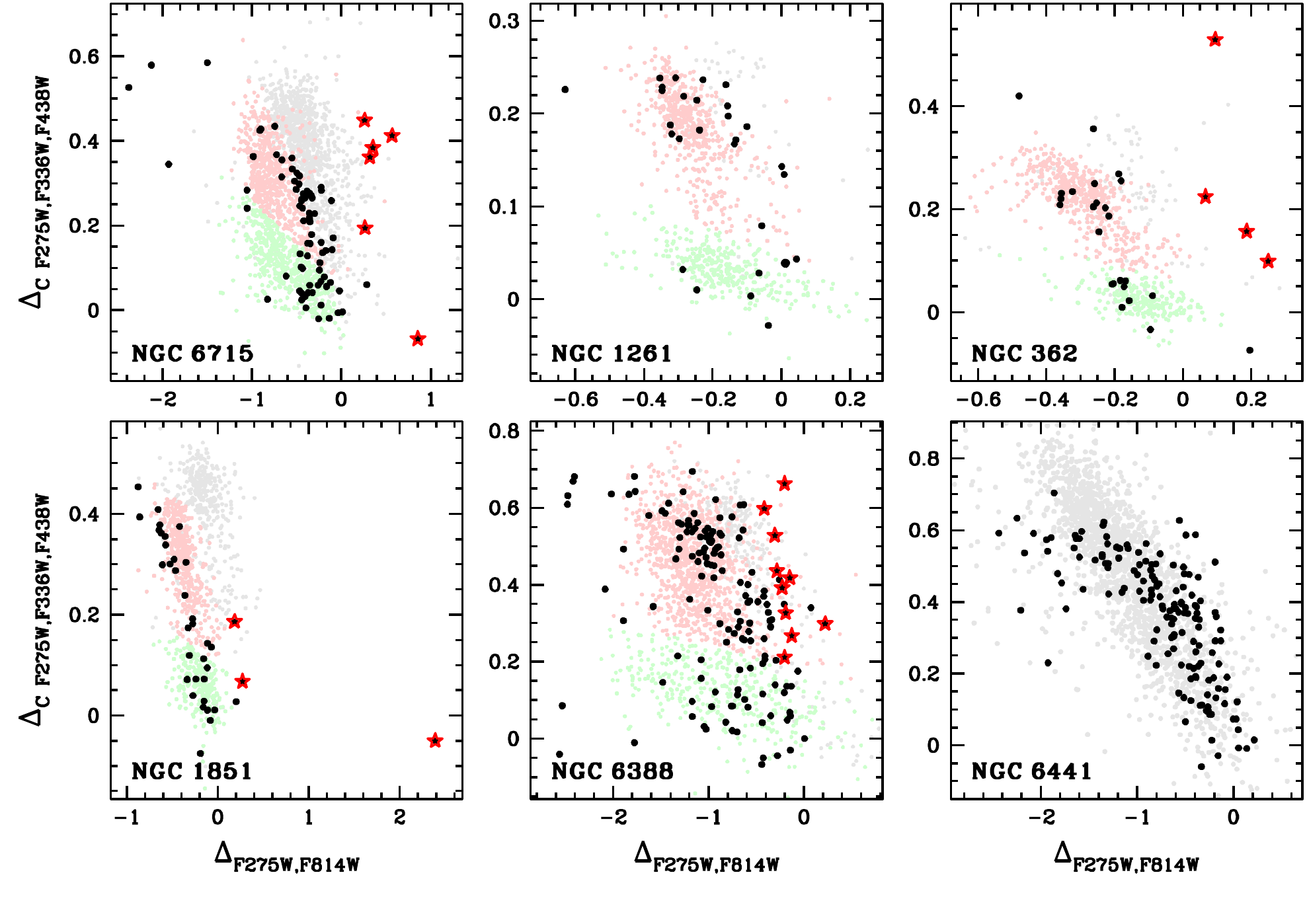} 
\caption{Chromosome map of the AGB stars of \ngc7078, \ngc2419, \ngc5024, \ngc6093,
	\ngc5286, \ngc7089, \ngc5986, \ngc6715, \ngc1261, \ngc362, \ngc1851,
	\ngc6388, \ngc6441, and \ngc5139. In each panel, AGB stars are
	represented as black points, and anomalous AGB stars are marked with a
	red starred symbol; light-green, magenta, and grey points indicate,
	respectively, the 1G, 2G, and anomalous RGB stars in the chromosome map
	of the corresponding cluster.  Since no univocal classification is
	available for \ngc6441 and \ngc5139, all the RGB stars of these two
	clusters are represented as grey points.  \label{fig:chms}}
\end{figure*}

We ended up with a sample of 14 GCs including \ngc362, \ngc1261, \ngc1851,
\ngc2419, \ngc5024, \ngc5139, \ngc5286, \ngc5986, \ngc6093, \ngc6388, \ngc6441,
\ngc6715, \ngc7078\, and \ngc7089. The ChMs of these GCs, sorted by increasing
metallicity, are displayed in Figure~\ref{fig:chms}. In the ChMs of type\,II
GCs we also highlighted the anomalous AGB stars with the usual red starred
symbol.  As for \ngc2808, in each panel we plotted on background the scaled-up
ChM of the corresponding cluster RGB stars. In passing we note that we have
also derived the ChM of the RGB stars of \ngc2419 for the first time in this
paper. For the sake of comparison, the probable 1G and 2G RGB members of each
cluster \citet{milone17} have been colored, respectively, in light-green and
magenta, except for \ngc2419 and \ngc6441, for which it is not possible to
obtain a straightforward population classification of their RGB stars.
In all the type\,II GCs diagrams, the anomalous RGB stars are represented
as grey points. 
 
In the majority of GCs, the distribution of the AGB stars does not appear
clustered in separate groups, except for \ngc7078, \ngc1261, \ngc362 and
perhaps \ngc6388, where we can recognize a subdivision that resembles that of
the overlapped RGB ChM. We also notice that, similarly to what is observed in
the RGB ChM, the anomalous AGB stars in type\,II GCs lie on the rightmost part
of the diagram.

The ChMs of Fig.~\ref{fig:chms} clearly show that, in each GC, the observed color spread
is not compatible with the presence of a single population. Moreover they show that the 
MPs in the AGB can be associated to the different groups visible in the RGB ChMs.

%%%%%%%%%%%%%%%%%%%%%%%%% FIGURE 16 %%%%%%%%%%%%%%%%%%%%%%%%%%%
%%%%%%%%%%%%%%%%%%%%%%%%%%%%%%%%%%%%%%%%%%%%%%%%%%%%%%%%%%%%%%%

\subsection{Population ratios}\label{sec:pop_r}
A quantitative estimate of the fraction of stars that skipped the AGB phase can
be obtained from the comparison of the ration between 2G stars in the AGB and
RGB. To identify the different groups of stars in the AGB we decided to adopt
the subdivision of the RGB ChM as a guideline for membership definition: all
the AGB stars falling in the area occupied by the RGB 1G (2G) stars have then
been flagged as 1G (2G) in the AGB. In type\,II GCs, the anomalous AGB stars
have been considered as part of the 2G sample.

This crude selection allowed us to estimate the population ratio
$\mathrm{(N_{1G}/N_{tot})_{AGB}}$ of the 1G AGB stars of \ngc2808 and 12 out of
14 GCs displayed in Fig.~\ref{fig:chms}. Indeed, as mentioned in the previous
section, it is not possible to derive a clear subdivisions of the RGB
populations of \ngc2419 and \ngc6441 from their ChMs. The resulting 1G
fractions, $\mathrm{(N_{1G}/N_{tot})_{AGB}}$, are listed in Table~\ref{tab2}.
The error of each observed population ratio has been obtained with the same
method used for the estimate of the anomalous AGB fraction, described in
Sect.~\ref{sec:anomAGB}. We see that the values range from $\sim 0.2$ in
\ngc5139 and \ngc5286 to $\sim 0.7$ in \ngc7078.  For comparison, we have also
reported the fraction of 1G RGB stars, $\mathrm{(N_{1G}/N_{tot})_{RGB}}$, from
\citet[][see their Table~2]{milone17}. 

\begin{figure*}
\centering
\includegraphics[width=\textwidth]{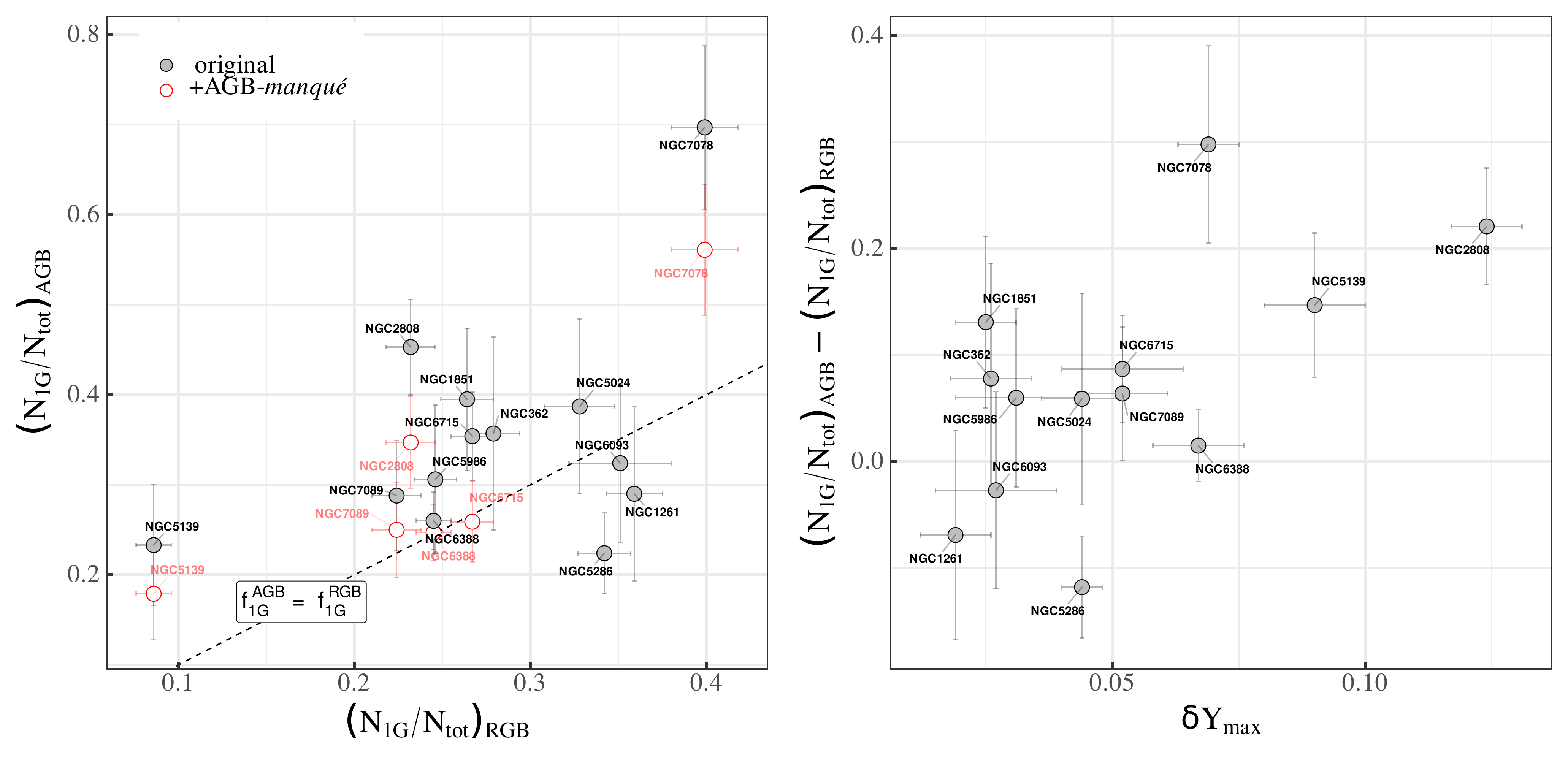} 
\caption{\textit{Left}: comparison between fraction of 1G AGB stars and 1G RGB
	stars \citep[from][]{milone17} for the 13 GCs in Table~\ref{tab2}. The
	dashed line marks the relation $\mathrm{(N_{1G}/N_{tot})_{AGB} =
	(N_{1G}/N_{tot})_{RGB}}$. The red points represent the location that
	\ngc2808, \ngc5139, \ngc6388, \ngc6715, \ngc7078, and \ngc7089 would
	have if AGB-\textit{manqu\'e} stars were included in the total AGB
	cluster population (see text for details). \textit{Right}: Difference
	between fraction of 1G AGB and RGB stars vs. maximum internal helium
	variation $\mathrm{\delta Y_{max}}$.\label{fig:f1G}}
\end{figure*}

In the left panel of Figure~\ref{fig:f1G} we plot the
$\mathrm{(N_{1G}/N_{tot})_{AGB}}$ vs $\mathrm{(N_{1G}/N_{tot})_{RGB}}$ of the
clusters in Table~\ref{tab2}. For reference, we also plotted a dashed line
marking the identity relation. We observe that four GCs, namely \ngc7089,
\ngc6388, \ngc1261, and \ngc6093 are consistent, within the errors, with an
equal proportion 1G stars in both the evolutionary phases. Four clusters,
namely \ngc5139, \ngc2808, \ngc6715, and \ngc7078, definitely exhibit a
shortage of 2G AGB stars: all these GCs have, indeed, a very prominent hot HB
component, expected to skip the EAGB and evolve instead along the
AGB-\textit{manqu\'e} channel. We also see that \ngc5286 has an higher fraction
of 1G RGB stars.

In the right panel of Fig.~\ref{fig:f1G} we plot the difference between the
fraction of 1G AGB and RGB stars as a function of the maximum internal helium
enrichment, $\mathrm{\delta Y_{max}}$, determined by \citet{milone18b}. We
observe a clear direct correlation between the two quantities. The trend
suggests that in the GCs with the most helium-enriched 2G component, the
fraction of 1G AGB stars is higher. These results represent a further,
independent confirmation of the predictions of the AGB-\textit{manqu\'e}
scenario.

%%%%%%%%%%%%%%%%%%%%%%%%% FIGURE 17 %%%%%%%%%%%%%%%%%%%%%%%%%%%
%%%%%%%%%%%%%%%%%%%%%%%%%%%%%%%%%%%%%%%%%%%%%%%%%%%%%%%%%%%%%%%

\begin{figure*}
\centering
\includegraphics[width=\columnwidth,trim=.7cm 4cm .2cm 4.5cm,clip]{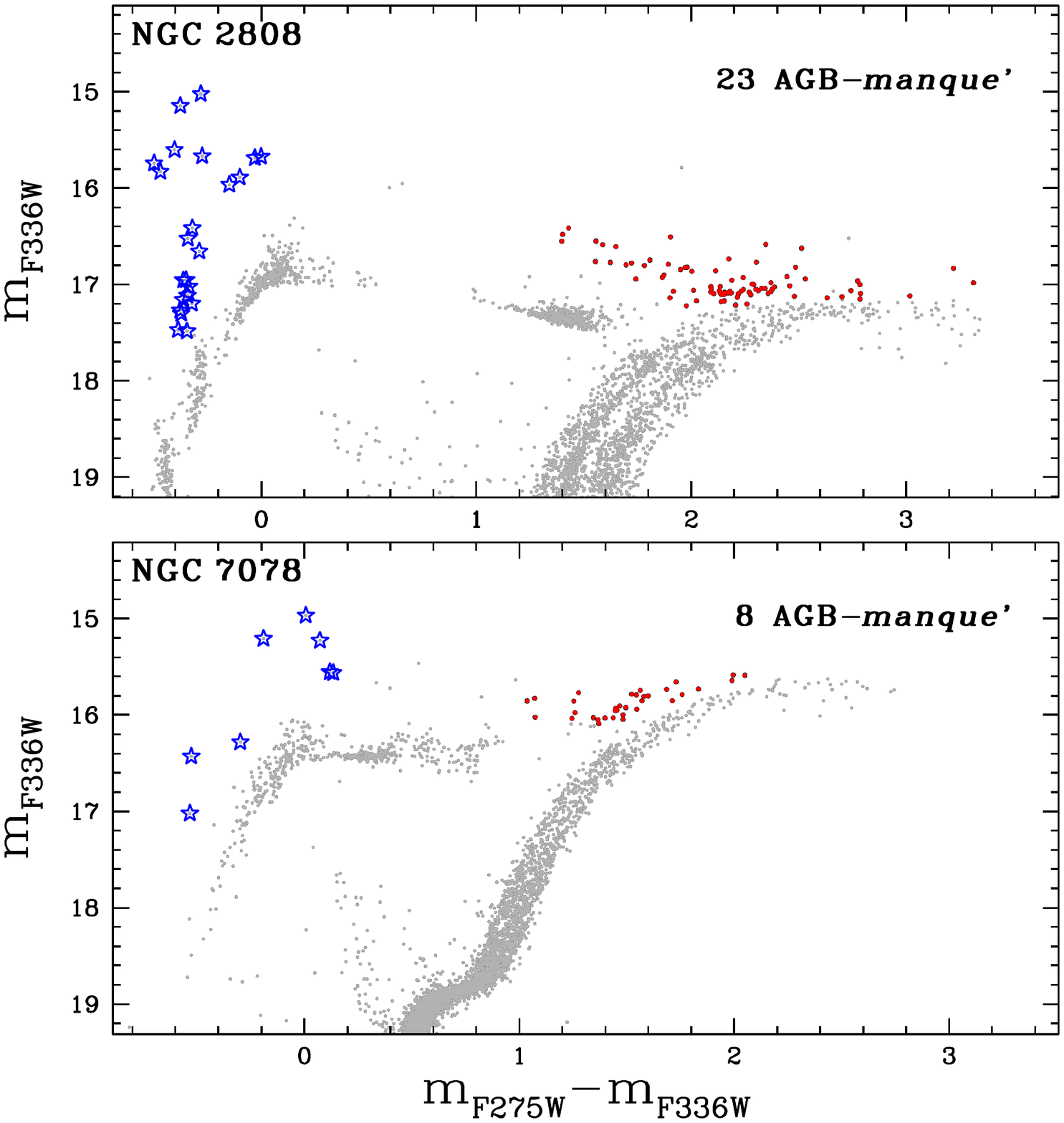} 
\includegraphics[width=\columnwidth,trim=.7cm 4cm .2cm 4.5cm,clip]{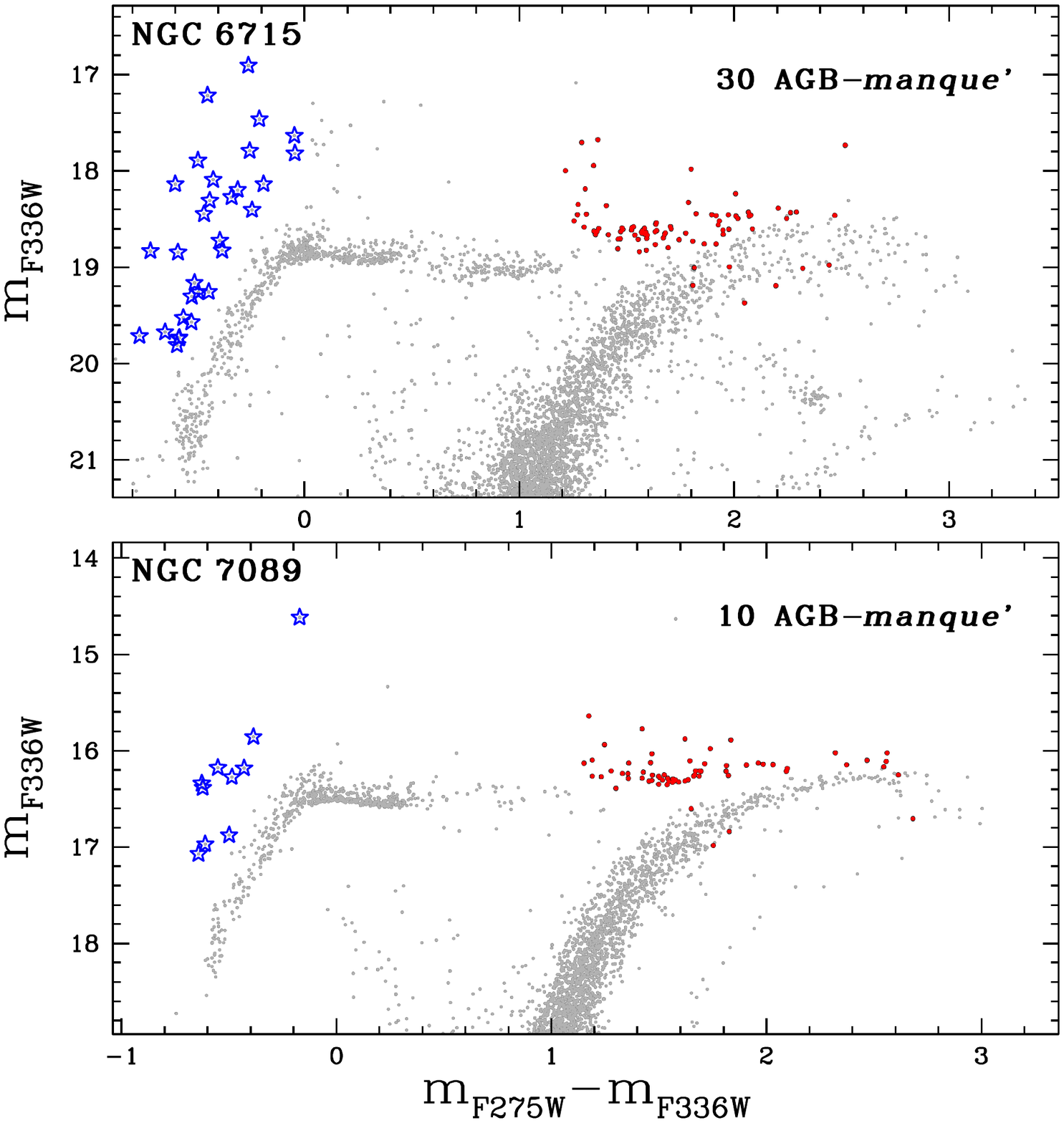} 
\caption{Identification of AGB-\textit{manqu\'e} candidates in four GCs with extended HB. 
	AGB stars are represented as red points, while AGB-\textit{manqu\'e} 
	candidates are marked with a blue starred symbol.\label{fig:agbm}}
\end{figure*}

As a simple experiment, we decided to verify to which extent the differences
observed between 1G AGB and RGB population ratios can be reconciled if the
stars evolved as AGB-\textit{manqu\'e} are included in the 2G population of the
AGB. To this aim, we sought AGB-\textit{manqu\'e} candidates in the
analyzed GCs by looking at their UV CMDs. An example is shown in
Figure~\ref{fig:agbm}, that displays the $m_{\rm F336W}$ vs. $m_{\rm
F275W}-m_{\rm F336W}$ CMD of four GCs with extended HB morphology: \ngc2808,
\ngc7078, \ngc6715, and \ngc7089. The AGB members of each cluster have been
represented as red points.

Since AGB-\textit{manqu\'e} stars have very thin envelopes, their evolution
will be entirely spent at \teff\ higher than $\approx 25,000$\,K, approaching
luminosities typical of post-EAGB stars \citep{greggio90}. According to this,
we expect to observe AGB-\textit{manqu\'e} stars to lie at the bluest end of
the HB, with luminosities comparable to those EAGB stars. Therefore, we
empirically selected as AGB-\textit{manqu\'e} candidates the stars occupying
the bluest portion of the CMD, with F336W magnitudes brighter than the group of
stars at the hottest end of the HB, which are mainly composed by EHB stars
\citep[e.g.][]{tailo15}. We marked the candidate stars with a blue starred
symbol. We cannot exclude the possible presence of few bright Blue Stragglers
in the selected samples.  Moreover, they could also potentially include
post-EAGB and/or post-AGB stars, although the probability to observe one of
these stars in a cluster is very low, because of their very rapid evolution
timescale ($\lesssim 10^6$\,yr) \citep{dcruz96,schiavon12a}. The low incidence
of these stars, however, does not affect our conclusions.

We identified 23 AGB-\textit{manqu\'e} candidates in \ngc2808, 9 in \ngc5139, 8
in \ngc6388, 30 in \ngc6715, 8 in \ngc7078, and 10 in \ngc7089. If the
AGB-\textit{manqu\'{e}} candidates represented the total fraction of lost stars
of the 2G population of these GCs, we would then observe a smaller fraction of
1G AGB stars, that is $\approx 0.35$ for \ngc2808, $\sim 0.18$ for \ngc5139,
$\sim 0.25$ for \ngc6388 and \ngc6715, $\sim 0.56$ for \ngc7078, and $\sim
0.25$ for \ngc7089, and relative errors comparable with the previous ones.  

The red points in the left panel of Fig.~\ref{fig:f1G} mark the new location in
the plot of the previous six GCs. We see that, in this case, \ngc6388,
\ngc6715, and \ngc7089 would be consistent, within one sigma, with the same
fraction of 1G AGB and RGB stars. We also notice that the fraction of 1G AGB
stars in \ngc6715 is slightly higher than the RGB one, and this is probably due
to some spurious additional contribution in the AGB-\textit{manqu\'e} count if
this cluster, which has a strong contamination from the background stars of the
Sagittarius Dwarf spheroidal galaxy. On the other hand, the new population
ratios of \ngc5139, \ngc2808 and \ngc7078 would be consistent with the same
fraction of 1G RGB stars only within about two sigma. We notice that since in
both the AGB-\textit{manqu\'e} and EAGB phase stars are burning helium in a
shell, the shell helium burning should run for the same time, therefore by
adding the AGB-\textit{manqu\'e} candidates, the AGB ratio should become equal
to the RGB ratio. While the fact that the anomalous population of \ngc5139
spans an extreme interval of metallicity \citep[$\sim 1.5$\,dex;
e.g.][]{johnson09,marino11a}, could potentially mitigate the observed
discrepancy in this cluster, the relative predominance of 1G AGB stars in the
other two GCs could be reconciled either by assuming a shorter evolutionary
timescale of the AGB-\textit{manqu\'e} stars. Indeed, since the major
difference between AGB-\textit{manqu\'e} and normal AGB stars is their core
mass within the hydrogen-helium discontinuity, the shorter timescale of the
former would derive from the fact that these stars are the progeny of hot
(extreme) HB stars, that experience almost no core-mass increase, as opposed to
the precursors of standard AGBs, namely normal HB stars, which instead undergo
a steady increase of the core mass \citep{greggio90}.

%%%%%%%%%%%%%%%%%%%%%%%%% FIGURE 18 %%%%%%%%%%%%%%%%%%%%%%%%%%%
%%%%%%%%%%%%%%%%%%%%%%%%%%%%%%%%%%%%%%%%%%%%%%%%%%%%%%%%%%%%%%%

%%%%%%%%%%%%%%%%%%%%%%%%%%%%%%%%%%%%%%%%%%%% TABLE 2 %%%%%%%%%%%%%%%%%%%%%%%%%%%%%%%%%%%%%%%%%%%%%%%%
\begin{deluxetable}{l*{2}{c}*{2}{C}}
\tabletypesize{\scriptsize}
\tablewidth{\columnwidth}
\tablecaption{Number of 1G and 2G AGB stars identified in 13 GCs, fraction of 1G AGB stars with respect to the total and corresponding fraction obtained from the RGB ChMs.\label{tab2}}
\tablehead{
\colhead{ID} & \colhead{$\mathrm{N_{1G}^{AGB}}$} & \colhead{$\mathrm{N_{2G}^{AGB}}$} & \multicolumn1c{$\mathrm{(N_{1G}/N_{tot})_{AGB}}$} & \multicolumn1c{$\mathrm{(N_{1G}/N_{tot})_{RGB}}$\tablenotemark{\footnotesize{a}}}
}
\decimals
\startdata
\ngc362  & 10 &  18 & 0.357 \pm 0.107 & 0.279 \pm 0.015 \\
\ngc1261 &  9 &  22 & 0.290 \pm 0.097 & 0.359 \pm 0.016 \\
\ngc1851 & 15 &  23 & 0.395 \pm 0.079 & 0.264 \pm 0.015 \\
\ngc2808 & 34 &  41 & 0.453 \pm 0.053 & 0.232 \pm 0.014 \\
\ngc5024 & 12 &  19 & 0.387 \pm 0.097 & 0.328 \pm 0.020 \\
\ngc5139 &  7 &  23 & 0.233 \pm 0.067 & 0.086 \pm 0.010 \\
\ngc5286 & 15 &  52 & 0.224 \pm 0.045 & 0.342 \pm 0.015 \\
\ngc5986 & 11 &  25 & 0.306 \pm 0.083 & 0.246 \pm 0.012 \\
\ngc6093 & 11 &  23 & 0.324 \pm 0.088 & 0.351 \pm 0.029 \\
\ngc6388 & 40 & 114 & 0.260 \pm 0.032 & 0.245 \pm 0.010 \\
\ngc6715 & 29 &  53 & 0.354 \pm 0.049 & 0.267 \pm 0.012 \\
\ngc7078 & 23 &  10 & 0.697 \pm 0.091 & 0.399 \pm 0.019 \\
\ngc7089 & 19 &  47 & 0.288 \pm 0.061 & 0.224 \pm 0.014 \\
\enddata
\tablenotetext{a}{From \citet{milone17}.}
\end{deluxetable}
%%%%%%%%%%%%%%%%%%%%%%%%%%%%%%%%%%%%%%%%%%%%%%%%%%%%%%%%%%%%%%%%%%%%%%%%%%%%%%%%%%%%%%%%%%%%%%%%%%%%%%%%%%%%%%%%%%

\section{AGB frequency and multiple populations}\label{sec:frequency}
The study of the AGB frequency relative to HB stars plays an important role in
the context of MPs. In particular, since AGB-\textit{manq\'e} stars are the
progeny of HB stars with mass lower than $\sim 0.5$\msun
\citep{moehler19,prabhu20}, which are associated with the most helium-rich
stellar component in a cluster \citep{dantona02a,tailo20}, differences in
internal helium variations between clusters can affect the AGB/HB ratio. This
prediction is quantified by the evolutionary parameter $\mathrm{R_2}$
\citep{caputo89}, namely the number ratio between AGB and HB stars, which is
proportional to the ratio between the lifetime of stars in these two
evolutionary phases. According to theoretical predictions, $\mathrm{R_2}$
should not be higher than about 0.2 \citep{cassisi14a}.

To explore this phenomenon, we took advantage of a large database of
homogeneously defined AGB samples in GCs, determined in the present work, and
of the independent measurements of maximum internal helium variations,
$\mathrm{\delta Y_{max}}$, obtained by \citep{milone18b}. For each GC, we
determined the frequency of AGB stars relative to HB stars,
$\mathrm{(N_{AGB}/N_{HB})}$, by counting HB members in the corresponding
$m_{\rm F606W}$ vs. $m_{\rm F606W}-m_{\rm F814W}$ CMD. The error of each
frequency measurement has been estimated with the same method used in
Section~\ref{sec:anomAGB}). With the purpose of limiting spurious outcomes due
to the stochastic fluctuations, we decided to limit the following analysis to
the 36 GCs in our database with $\mathrm{N_{AGB}} \geq 10$, for which we report
the corresponding frequency values Table~\ref{tab3}.

%%%%%%%%%%%%%%%%%%%%%%%%%%%%%%%%%%%%%%%%%%%% TABLE 3 %%%%%%%%%%%%%%%%%%%%%%%%%%%%%%%%%%%%%%%%%%%%%%%%
\begin{deluxetable}{lC}
\def\arraystretch{0.95}
\tablewidth{\columnwidth}
\tablecaption{Observed AGB/HB frequency for the clusters in our database with $\mathrm{N_{AGB}} \geq 10$.\label{tab3}}
\tablehead{
\colhead{ID} & \multicolumn1c{$\mathrm{N_{AGB}/N_{HB}}$} 
}
\decimals
\startdata
\ngc104  & 0.099 \pm 0.015 \\ 
\ngc362  & 0.086 \pm 0.014 \\
\ngc1261 & 0.161 \pm 0.022 \\
\ngc1851 & 0.126 \pm 0.018 \\
\ngc2419 & 0.185 \pm 0.017 \\
\ngc2808 & 0.108 \pm 0.011 \\
\ngc4833 & 0.117 \pm 0.026 \\
\ngc5024 & 0.111 \pm 0.019 \\
\ngc5139 & 0.172 \pm 0.024 \\
\ngc5272 & 0.089 \pm 0.022 \\
\ngc5286 & 0.172 \pm 0.017 \\
\ngc5904 & 0.105 \pm 0.025 \\
\ngc5927 & 0.071 \pm 0.016 \\
\ngc5986 & 0.124 \pm 0.017 \\
\ngc6093 & 0.125 \pm 0.019 \\
\ngc6101 & 0.151 \pm 0.036 \\
\ngc6171 & 0.289 \pm 0.052 \\
\ngc6205 & 0.058 \pm 0.016 \\
\ngc6254 & 0.132 \pm 0.035 \\
\ngc6304 & 0.092 \pm 0.023 \\
\ngc6341 & 0.114 \pm 0.027 \\
\ngc6388 & 0.175 \pm 0.011 \\
\ngc6441 & 0.120 \pm 0.009 \\
\ngc6584 & 0.125 \pm 0.030 \\
\ngc6624 & 0.106 \pm 0.026 \\
\ngc6637 & 0.158 \pm 0.024 \\
\ngc6656 & 0.167 \pm 0.043 \\
\ngc6681 & 0.119 \pm 0.027 \\
\ngc6715 & 0.128 \pm 0.012 \\
\ngc6723 & 0.149 \pm 0.028 \\
\ngc6752 & 0.092 \pm 0.023 \\
\ngc6779 & 0.127 \pm 0.032 \\
\ngc6934 & 0.176 \pm 0.032 \\
\ngc6981 & 0.179 \pm 0.043 \\
\ngc7078 & 0.077 \pm 0.013 \\
\ngc7089 & 0.126 \pm 0.013 \\
\enddata
\end{deluxetable}
%%%%%%%%%%%%%%%%%%%%%%%%%%%%%%%%%%%%%%%%%%%%%%%%%%%%%%%%%%%%%%%%%%%%%%%%%%%%%%%%%%%%%%%%%%%%%%%%%%%%%%%%%%%%%%%%%%

\begin{figure*}
\centering
\includegraphics[width=\textwidth]{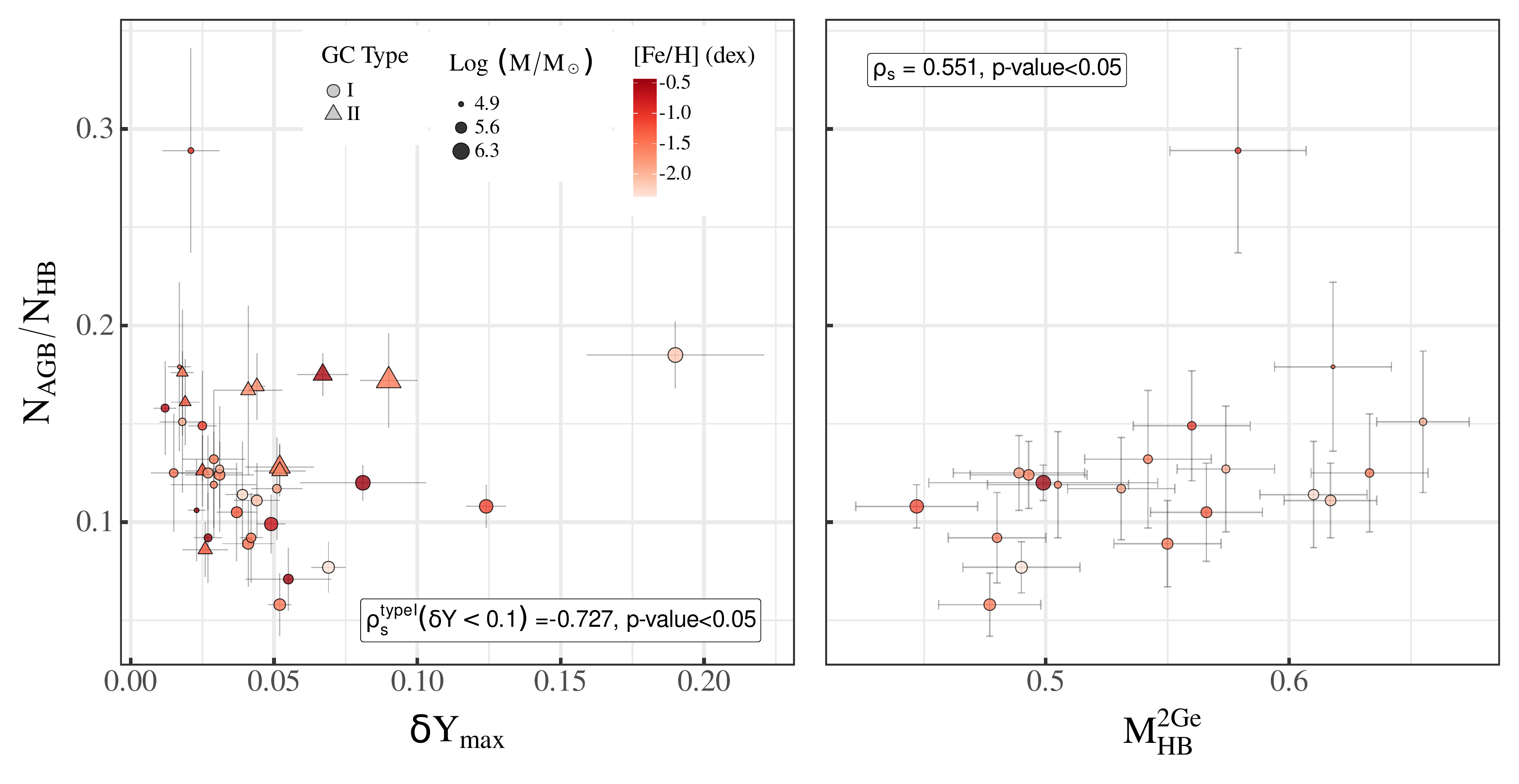} 
\caption{\textit{Left}:AGB/HB frequency stars vs.
	maximum internal helium variation $\mathrm{\delta Y_{max}}$. Shape, size
	and color of each points indicate, respectively, type, total mass and
	metallicity of the corresponding cluster. \textit{Right}: 
	AGB/HB frequency vs. average mass of the most
	helium-enriched population of 20 type\,I GCs in our database. 
	\label{fig:fAGB-HB}}
\end{figure*}

In the left panel of Figure~\ref{fig:fAGB-HB} we plot of $\mathrm{(N_{AGB}/N_{HB})}$
and ${\rm \delta Y_{max}}$ for these GCs, where color, size and shape of each
point map, respectively, metallicity \citep[][2010 update]{harris96a}, present-day mass
\citep{baumgardt18b}, and cluster's type \citep{marino19}, as indicated in the
legend. The diagram shows a scattered distribution, with all the GCs having
frequency values smaller than $\sim 0.2$ except \ngc6171, which has
$\mathrm{(N_{AGB}/N_{HB})} \lesssim 0.3$. We observe that the bulk of GCs does
not seem to follow any clear trend. It is necessary to consider that the AGB of
type\,II GCs can suffer a significant depletion of anomalous population stars,
as seen in Section~\ref{sec:anomAGB}. Hence, the resulting shortage of AGB
stars in these GCs would importantly affect the observed AGB/HB frequency meaning
in a way difficult to predict. This implies in turn that no strong conclusions
can be derived for these GCs.

On the other side, if we limit our analysis to type\,I GCs with
$\log (M/M_{\odot})\lesssim 6$, therefore excluding the two massive GCs \ngc2808,
\ngc2419, which are the type\,I clusters with the highest internal helium
variation \citep[$\mathrm{\delta Y_{max}}>0.1$][]{milone18b}, and among the GCs
with the most complex MP pattern \citep{milone15b,zennaro19}, the majority of
points display an inverse monotonic trend, with smaller AGB frequencies
attained by clusters with higher internal helium variations. This impression is
confirmed by the high value of the Spearman's coefficient ($\rho_{\rm s} =
-0.727$) reported in the bottom right corner.  Finally we notice that the
distribution of points shows no correlation with the cluster metallicity.

Finally, we verified that the analysis of the AGB/RGB frequency shows a
similar trend, resulting therefore in the same conclusions. For the sake of
convenience we decided to not show the corresponding plot.

%%%%%%%%%%%%%%%%%%%%%%%%% FIGURE 19 %%%%%%%%%%%%%%%%%%%%%%%%%%%
%%%%%%%%%%%%%%%%%%%%%%%%%%%%%%%%%%%%%%%%%%%%%%%%%%%%%%%%%%%%%%%

Another way of investigating AGB depletion and helium enrichment is through the
analysis of the relation between observed frequency and minimum mass along the
HB. \citet{gratton10} showed that a direct correlation exists between minimum
mass along the HB and the numeric frequency of AGB stars relative to the RGB,
with lower frequencies attained by clusters with smaller minimum HB mass.  In
the right panel of Figure~\ref{fig:fAGB-HB} we plot $\mathrm{(N_{AGB}/N_{HB})}$ vs
$\mathrm{M_{HB}^{2Ge}}$, where the latter quantity indicates the average
mass of the HB stars belonging to the most helium-enriched population, recently
determined by \citet{tailo20}, available for 20 type\,I GCs in our database.
Color and size of each points follows the convention adopted in the left panel.
With the exception of the outlier \ngc6171, we observe a clear correlation,
that confirms the expectation of the AGB-\textit{manq\'{e}} scenario of lower
AGB frequencies in clusters with smaller minimum mass in the HB
\citep{prabhu20}.  The mild monotonic trend is also indicated by the value of
the Spearman's coefficient, $\rho_{\rm s} = 0.551$,
In passing we also notice that clusters with higher total mass have, on
average, smaller minimum HB mass, while no clear trend is visible with
metallicity. 

The two plots in Figure~\ref{fig:fAGB-HB} indicate, in general, a mild correlation
between frequency of AGB stars and internal helium variations in clusters,
which is in turn directly related to the mass of their hottest HB stars.  The
lack of a strong evidence of correlation is potentially connected to the fact
that the parameters ruling the distribution of HB stars are degenerate. For
instance, at a given helium enrichment, we would expect more massive HB stars
in slightly younger GCs, thus a smaller proportion of \textit{AGB-manqu\'e}.
The same conclusion would also be true for clusters of the same age, but different
metallicity.

\section{Summary and conclusions\label{sec:end}}
MPs are a common property of Galactic GCs. They have been widely characterized
along main sequence, RGB and HB stars in large samples of clusters
\citep[e.g.][]{anderson97,dantona05,piotto07,milone17,lagioia19,marino19,dondoglio21,tailo20}.
A coherent picture at later evolutionary stages is, however, still missing.

Recent spectroscopic analyses revealed that a lower fraction of 2G stars is
found to populate the AGB of some clusters, like \ngc6752
\citep{campbell13,lapenna16}, \ngc6121 \citep[M\,4]{marino17,maclean16},
\ngc6205 \citep[M\,62]{lapenna15a} and \ngc2808 \citep{marino17}. Although no
univocal consensus has been achieved on the interpretation of the results, the
lack of a more or less significant fraction of 2G stars corresponds to the
predictions of the AGB-\textit{manqu\'{e}} scenario \citep{greggio90},
according to which the most helium-enriched stars have envelope masses too low
in the HB phase, so they do not evolve towards the Hayashi track as a normal
AGB star, but rather towards higher temperatures until they reach the
white-dwarf cooling sequence.

In order to shed new light to this problem we decided to analyze the
photometric properties of the AGB stars in the largest sample of GCs analyzed
so far, composed by 58 objects observed by \hst\ in UV and optical bands
\citep{sarajedini07a,piotto15}. Suitable combinations of filters allow to
disentangle MPs in CMDs and study their properties in statistically significant
samples of stars. Through a procedure that exploits CMDs in different color
combinations we selected the most probable AGB candidates in every analyzed
cluster.  Then, we derived their properties through the analysis of their
photometric features and comparison with theoretical models.  Our methods and
findings are summarized in the following: 
\begin{enumerate}

\item we investigated AGB stars in type\,II GCs and identified, for the first
	time, AGB stars associated with their metal-rich (anomalous) stellar
		populations. The fraction of anomalous AGB stars with
		respect to the total number of AGB stars is significantly lower
		than the corresponding fraction of stars in the anomalous RGB, with
		\ngc7089 and \ngc362 possible exceptions;  

\item we constructed the $m_{\rm F814W}$ vs. ${C_{\rm F275W,F336W,F438W}}$ CMD
	of the bright stars of 56 GCs with at least one AGB detection. We find
		that the AGB sequences of at least 48 clusters
		exhibit larger pseudo-color spread than that expected from
		photometric errors alone. Hence, they host multiple stellar
		populations. In eight clusters, namely \ngc7099, \ngc4590,
		\ngc5466, \ngc6218, \ngc6397, \ngc6496, \ngc6535, and \ngc6809,
		there is no evidence of intrinsic ${C_{\rm F275W,F336W,F438W}}$
		broadening among AGB stars with similar luminosity. However,
		the small number of AGB stars in these clusters prevents us
		from any conclusion on the presence of MPs among their AGB. We
		conclude that in none of the studied clusters all 2G stars
		avoid the AGB phase;

\item we used synthetic spectra with appropriate chemical composition and
	theoretical stellar models from the Roma database
		\citep{ventura98a,mazzitelli99} to derive, for the first time,
		isochrones of 1G and 2G AGB stars. We show that 2G stars define
		AGB sequences with smaller values of $C_{\rm
		F275W,F336W,F438W}$ and bluer ${M_{\rm F275W}-M_{\rm F814W}}$
		colors than 1G stars with the same luminosity. The flux
		difference is mostly due to NH molecular bands that affect the
		spectral region covered by the F336W filter and the OH and CN
		molecules enclosed by the F275W and F438W filters,
		respectively.  Since 2G stars are enhanced in N and depleted in
		C and O, they have fainter F336W and brighter F275W and F438W
		magnitudes than 1G stars with the same luminosity. Moreover, 2G
		stars are helium enhanced with respect to the 1G, hence they
		have colors bluer than 1G stars at the same luminosity. In
		addition to the content of helium, C, N, and O, the amount of
		mass loss experienced during the RGB phase affects the colors
		of AGB stars. In particular, an increase in mass loss can
		dramatically shift to the blue the AGB sequences;

\item we measured the intrinsic width of AGB stars in the pseudo-colors
	${C_{\rm F275W,F336W,F438W}}$ for 35 GCs with ${\mathrm
		N_{AGB}\geq10}$, and compared it with the corresponding
		quantity obtained from RGB stars. We observe that the intrinsic
		width of AGB stars is on average smaller than that of RGB.
		This finding suggests that a substantial part of the RGB stars in
		these GCs does not evolve to the AGB;

\item we constructed the ChM of 15 GCs with well populated AGB sequence and
	identified their 1G and 2G AGB stars: this allowed us to derive the
		fraction of 1G stars along the AGB. On average, AGB stars host
		fractions of 1G stars larger than in the RGB, and \ngc2808,
		\ngc5139, \ngc6715, and \ngc7078 and are the clusters with the
		most pronounced differences. This result is consistent with the
		AGB-\textit{manqu\'{e}} scenario where part of 2G stars skips
		the AGB phase \citep{campbell13} due to their typical mass lower than
		that of 1G stars. Indeed, 2G stars are enhanced in helium
		\citep{lagioia18,milone18b} and lose more mass in the RGB phase
		compared to the 1G \citep{tailo19a,tailo20}. This scenario is
		corroborated by the evidence from this paper and from
		\citet{gratton10} that the frequency of AGB/HB stars of the
		type\,I GCs correlates with the minimum mass of HB stars; the
		AGB frequency of type\,I GCs with $\log (M/M_{\odot})\lesssim
		6$ also anti-correlates with the maximum helium abundance of 2G
		stars. 

\end{enumerate}

\acknowledgments
This work has received funding from the European Research Council (ERC) under
the European Union's Horizon 2020 research innovation program (Grant
Agreement ERC-StG 2016, No 716082 `GALFOR', PI: Milone). APM, MT and ED have
been supported by MIUR under PRIN program 2017Z2HSMF
\facility{HST (ACS, WFC3)}

\bibliography{ms}

\end{document}